\documentclass[fleqn,usenatbib,twocolumn]{mnras}

\usepackage{newtxtext,newtxmath}

\usepackage[T1]{fontenc}
\usepackage{ae,aecompl}

\usepackage{etoolbox}
\makeatletter
\patchcmd\@combinedblfloats{\box\@outputbox}{\unvbox\@outputbox}{}{%
}%
\makeatother

\usepackage{graphicx}	
\usepackage{amsmath}	
\usepackage{float}
\usepackage{academicons}
\usepackage{xcolor}
\usepackage{orcidlink}
\usepackage{hyperref}
\usepackage{color,soul}  

\newcommand{\lnQ}{\ln Q_0}
\newcommand{\sigvar}{\sigma}

\usepackage[a4paper]{geometry}
\newcommand{\orcid}[1]{\href{https://orcid.org/#1}{\textcolor[HTML]{A6CE39}{\aiOrcid}}}

\title
[Particle acceleration in radio galaxies with flickering jets]
{
Particle acceleration in radio galaxies with flickering jets: GeV electrons to ultrahigh energy cosmic rays
}

\author[J.~H.~Matthews and A. M. Taylor]{James~H.~Matthews$^1$\thanks{matthews@ast.cam.ac.uk}\orcidlink{0000-0002-3493-7737} and Andrew~M. Taylor$^2$\orcidlink{0000-0001-9473-4758}
\\$^1$Institute of Astronomy, University of Cambridge, Madingley Road, Cambridge, CB3 0HA, UK\\
$^2$DESY, D-15738 Zeuthen, Germany
}

\date{Accepted 2021 March 10. Received 2021 March 9; in original form 2020 December 11}

\pubyear{2020}

\begin{document}
\label{firstpage}
\pagerange{\pageref{firstpage}--\pageref{lastpage}}
\maketitle

\begin{abstract}
Variability is a general property of accretion discs and their associated jets. We introduce a semi-analytic model for particle acceleration and radio jet/lobe evolution and explore the effect of Myr timescale jet variability on the particles accelerated by an AGN jet. Our work is motivated by the need for local powerful ultrahigh energy cosmic ray (UHECR) sources and evidence for variability in AGN and radio galaxies. Our main results are: i) UHECR and nonthermal radiative luminosities track the jet power but with a response set by the escape and cooling times, respectively; ii) jet variability produces structure in the electron, synchrotron and UHECR spectra that deviates from that produced for a constant jet power -- in particular, spectral hardening features may be signatures of variability; iii) the cutoff in the integrated CR spectrum is stretched out due to the variation in jet power (and, consequently, maximum CR energy). The resulting spectrum is the convolution of the jet power distribution and the source term. We derive an approximate form for a log-normal distribution of powers; iv) we introduce the idea of $\sim10\,$GeV ‘proxy electrons’ that are cooling at the same rate that UHECRs of rigidity $10\,$EV are escaping from the source, and  determine the corresponding photon frequencies that probe escaping UHECRs. Our results demonstrate the link between the history of an astrophysical particle accelerator and its particle contents, nonthermal emission and UHECR spectrum, with consequences for observations of radio galaxies and UHECR source models.
\end{abstract}


\def\bb{\boldsymbol}
\def\alfven{Alfv\'en}
\def\alfvenic{Alfv\'enic}
\newcommand{\ergs}{erg~s$^{-1}$}
\newcommand{\tauesc}{\tau_{\rm esc}^{\rm cr}}

\section{Introduction}
Multi-wavelength variability in the observed fluxes of accreting sources is observed on a range of timescales. This flux variability has a number of near-universal characteristics. Often, the variability is consistent with ``flicker'' noise, also known as $1/f$ noise or pink noise because the power spectrum follows a $1/f$ slope that is steeper than white noise but shallower than red. A pedagogical discussion of flicker noise in astronomy is given by \cite{press_flicker_1978}. A linear `rms-flux relation' has been observed in X-ray binaries \citep{uttley_flux-dependent_2001,uttley_non-linear_2005,heil_ubiquity_2012}, accreting white dwarfs \citep{scaringi_universal_2012,van_de_sande_rms-flux_2015}, blazars \citep{biteau_minijets---jet_2012} and other X-ray AGN \citep{uttley_non-linear_2005}. A related phenomenon is a log-normal distribution of observed fluxes, seen in both disc-dominated \citep{gaskell_lognormal_2004,alston_non-stationary_2019}  and jet-dominated sources (\citealt{h_e_s_s_collaboration_vhe_2010,h_e_s_s_collaboration_characterizing_2017,chevalier_variability_2019}; see also \citealt{morris2019}). In addition to flickering-type variability, sources also undergo episodic bursts of accretion activity, with prominent  examples being the outburst cycles of X-ray binaries \citep{fender_towards_2004}, restarting radio galaxies \citep{kaiser_radio_2000,brienza_duty_2018,konar_mode_2019} and, possibly, the optical `changing-look' phenomenon in quasars \citep{lamassa_discovery_2015,macleod_systematic_2016,runnoe_now_2016}. 

The accretion disc is intimately connected to the outflows produced by the system, which can take the form of winds or relativistic jets. If the jets are launched by the \cite{blandford_electromagnetic_1977} mechanism, then for a black hole (BH) with gravitational radius $r_g$, the power of the jet depends on the BH spin parameter ($a_*$) and magnetic flux ($\Phi_B$) threading the event horizon as $Q_{\rm BZ} \propto c (a_* \Phi_B / r_g)^2$. The precise relationship between the accretion rate, $\dot{m}$, and jet power is complicated; the production of powerful jets might require special conditions such as a magnetically arrested disk (MAD) state \citep{narayan_magnetically_2003,tchekhovskoy_efficient_2011,liska_large-scale_2018} and/or the presence of a disc wind to collimate the flow \citep[e.g.][]{globus2016,blandford_relativistic_2019}. If we assume that $\Phi_B^2$ is proportional to $(\dot{m} c r_g^2)$ and that $a_*$ changes on relatively long timescales, then $Q_{\rm BZ} \propto \dot{m} c^2$. This proportionality is also expected on general energetic grounds \citep[e.g.][]{chatterjee2019,davis2020}. It is therefore reasonable that the jet variability is in some sense a filtered version of the accretion-induced variability, a concept which has been successfully applied to jet modelling in X-ray binaries \citep{malzac_spectral_2014,malzac_jet_2018}. What effect does this jet variability have in AGN and radio galaxies? How do multiple episodes of accretion, which are themselves variable, affect the jet propagation, feedback and any observable radiative or UHECR signatures?

In AGN, there are also longer-term aspects of the accretion process that can imprint variability in the observed accretion flux and the power in the jet. For example, on long timescales ($\gtrsim$Myr), the accretion rate is likely to be determined by the supply of cold gas to the central region of the galaxy. In the chaotic cold accretion model  proposed by \cite{gaspari_self-regulated_2016}, the accretion rate on to the BH is predicted to follow a log-normal distribution with a pink/flicker noise power spectrum, just as is observed on shorter timescales in accretion discs. Simulations by \cite{yang_how_2016} also show flickering type variability in the jet power caused by self-regulated accretion on to the central AGN, with jet power consequently varying over a large range, \citep[$>2$ dex; see also][]{beckmann_dense_2019}. Long-term variability in jet power therefore seems inevitable from a fuelling perspective. 

Astrophysical jets accelerate nonthermal particles (see \citealt{matthews_particle_2020} for a review), which radiate as they interact with magnetic fields or radiation. Our primary way of learning about radio galaxies, is, as the name suggests, through radio emission from synchrotron-emitting electrons. As well as these nonthermal electrons, AGN jets can accelerate high-energy protons and ions, which we refer to as cosmic rays (CRs). The origin of the highest energy CRs, known as ultrahigh energy cosmic rays (UHECRs), is not known. The maximum energy attainable in a particle accelerator is given by the Hillas energy $E_H = Ze\beta BR$, where $\beta=u/c$ is the characteristic velocity, $B$ is the magnetic field strength and $R$ is the size of the acceleration region. The Hillas energy is a general constraint that can be understood in terms of moving a particle of charge $Ze$ a distance $R$ through an optimally arranged $-\bb{u}\times\bb{B}$ electric field. The Hillas criterion states that any accelerator must be a factor $\beta^{-1}$ larger than the Larmor radius of the highest energy particles it accelerates, i.e. $R>\beta^{-1} R_g(E_H)$, where $R_g$ denotes the Larmor radius. Calculating $E_H$ with some characteristic numbers for radio galaxies reveals they are one of the few sources capable of reaching $>$EeV energies and accelerating UHECRs. For this reason, they have long been discussed as potential UHECR sources \citep[e.g.][]{hillas_origin_1984,norman_origin_1995,hardcastle_high-energy_2009,eichmann_ultra-high-energy_2018,matthews_fornax_2018,matthews_cosmic_2019}, along with other classes of AGN jets and alternative sources such as gamma-ray bursts (GRBs), starburst winds and cluster-scale shocks. 

The Hillas energy can be used to obtain a power requirement for UHECR production, first derived using a dynamo model by \cite{lovelace_dynamo_1976}, and discussed further by various authors \citep[e.g.][]{waxman_cosmological_1995,blandford_acceleration_2000,nizamov_constraints_2018,eichmann_ultra-high-energy_2018}. This power requirement is sometimes referred to as the Lovelace-Hillas or magnetic luminosity condition. If a particle reaches the Hillas limit, then the magnetic power must satisfy $Q_B \gtrsim \beta^{-1}~10^{43}~\mathrm{erg~s}^{-1}$ for acceleration to a rigidity of $E/Ze=10$EV. For reasonable departures from equipartition and accounting for a likely maximum energy some factor below $E_H$, UHECR acceleration probably requires kinetic jet powers in the region $Q_{\rm j} \gtrsim 10^{44}$\ergs. There are relatively few nearby radio galaxies (within the various UHECR horizons) capable of reaching these powers, if the current jet power estimates are taken at face value \citep{massaglia_role_2007,massaglia_radio_2007,matthews_fornax_2018,matthews_cosmic_2019}.

Any variability in the jet power will inevitably affect the morphology and dynamics of radio galaxies, as well as any feedback processes at work. There are now many examples of restarting or `double-double' radio galaxies \citep[e.g.][]{kaiser_radio_2000,brienza_duty_2018,konar_mode_2019}, in which there appear to be distinct, discrete episodes of jet activity, but there is also subtler evidence for variability in the jet power. For example, Fornax A displays evidence for a varied and complex history \citep{iyomoto_declined_1998,lanz_constraining_2010,maccagni_flickering_2020}, while Centaurus A shows a distinction between different lobe structures on scales of $\sim2\,$kpc and $\sim300\,$kpc \citep{morganti_centaurus_1999,croston_high-energy_2009}, which appear to be connected with different episodes of activity. The merger and gas fuelling history of the sources is complex in both Fornax A \citep[e.g.][]{iyomoto_declined_1998,mackie_evolution_1998,iodice_fornax_2017} and Centaurus A \citep[e.g.][]{stickel_first_2004,neff_complex_2015}. In a more general sense, it is clear that there are many aspects of radio galaxies that are time integrated -- for example, the total energy input into the surroundings, or the total energy stored in synchrotron-emitting electrons -- and variability creates a disconnect between these integrated quantities and the instantaneous jet properties.

In this paper, our aims are threefold: (i) to introduce a numerical method capable of modelling, in a simple parameterised fashion, the morphology, radiation and UHECR signatures from radio galaxies with variable jet power; (ii) to study the effect of jet variability on observational signatures such as the synchrotron luminosity and broadband spectral energy distribution; (iii) to study the acceleration and escape of UHECRs in radio galaxies with variable jet powers. In our modelling, we focus on jets with a flicker noise power spectrum and a log-normal power distribution. We make a number of further simplifying assumptions and the model is unlikely to provide quantitative matches with real astrophysical sources. Indeed, our approach is mostly heuristic -- we aim to demonstrate some key principles regarding particle acceleration in variable jets that can be used to study UHECR and electron acceleration. We begin by describing our method (section 2) and present the results from a single simulation in section 3. In section 4, we introduce the concept of `proxy electrons', before discussing extragalactic CR propagation, observational applications and limitations of the model. We conclude in section 5.

\section{A Simple Variable Jet and Particle Injection Model}
\label{sec:method}
We begin by introducing our model for the evolution of a flickering jet and the nonthermal particles it accelerates. The jet has a kinetic power that behaves stochastically according to an input power spectrum and probability density function; we begin by describing the process of generating this jet power time series, before describing the dynamics of the jet and the treatement of particle acceleration. 

\subsection{Synthetic Jet Power Histories}
\label{sec:synthetic_jet}
\begin{figure}
    \centering
	\includegraphics[width=0.9\linewidth]{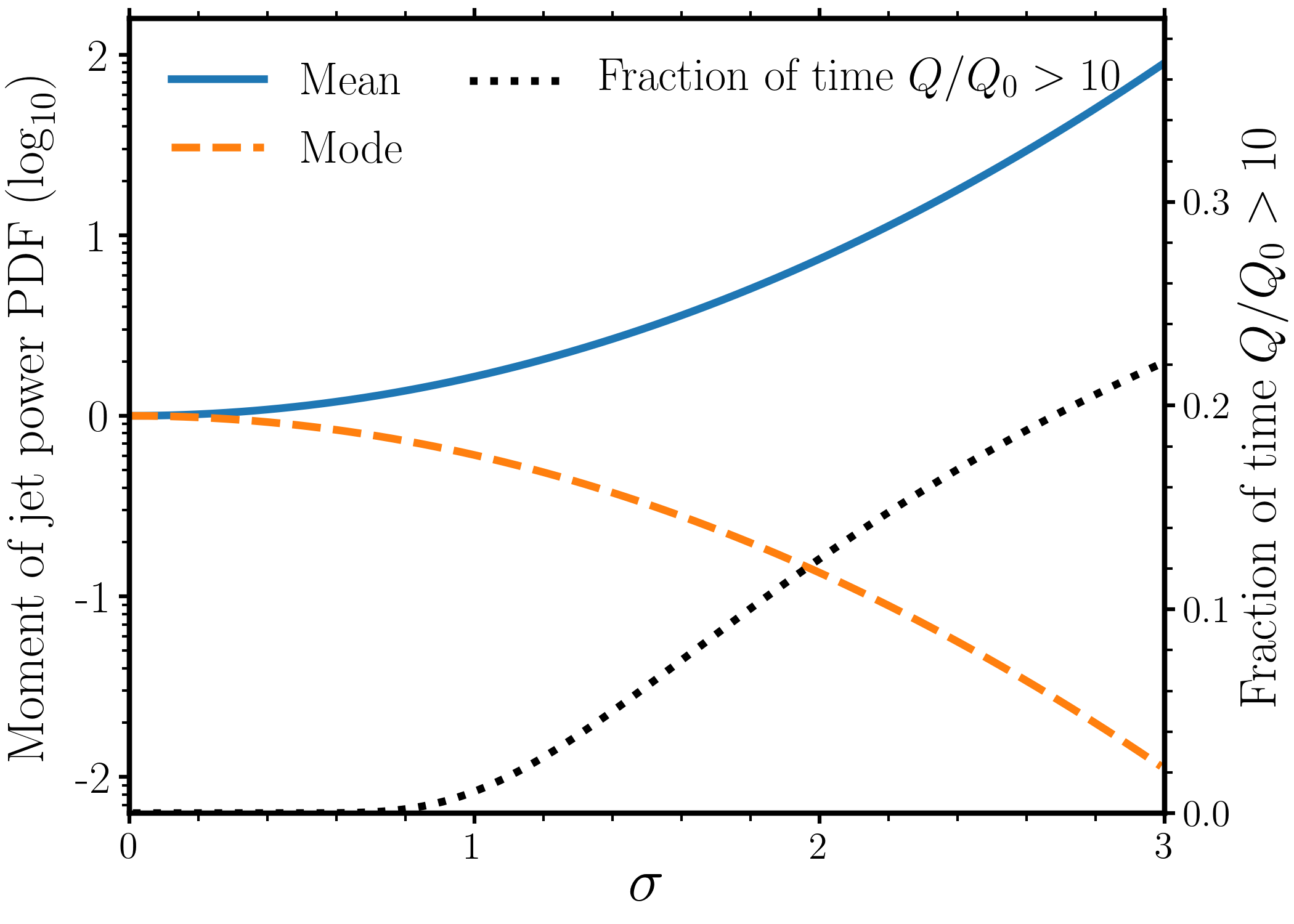}
    \caption{
    The relationship between various properties of a log-normal jet power distribution, plotted on a logarithmic (base 10) scale. The blue solid and orange dashed lines show the mean and mode of the distribution, given by $e^{\lnQ+\sigvar^2/2}$ and $e^{\lnQ-\sigvar^2}$ respectively. The mean increases as $\sigvar$, the natural logarithm of the standard deviation, increases. The black dotted line is shown on a different (linear) $y$-axis, and shows the fraction of time the jet has a power exceeding $10 Q_0$. This is equivalent to the survival function of the distribution (or one minus the CDF) evaluated at $10~Q_0$\ergs. The plot illustrates how a wider log-normal distribution (a more variable jet power) produces more favourable conditions for UHECR acceleration; as $\sigvar$ increases, the jet spends more time in a `high' state and has a higher integrated energy output. 
    }
    \label{fig:sigma}
\end{figure}

To create a synthetic jet power time-series, we use an input power spectral density (PSD) and jet power probability density function (PDF) to generate a time series for the jet power for a given set of PSD and PDF parameters. We use the method described by \cite{emmanoulopoulos_generating_2013} and implemented in python by \cite{connolly_python_2015}. The algorithm is similar to the widely used \cite{timmer_generating_1995} method, except that it allows for more flexibility in specifying the underlying 
The PSD is specified as a power-law model of the form 
\begin{equation}
    \mathrm{PSD}(f) \propto f^{-\alpha_{p}}
\end{equation}
where $f$ is the temporal frequency. We set $\alpha_{p}=1$ for a ``pink'' or flicker noise spectrum and use bins of $0.1$ Myr when generating the synthetic time series. We use a log-normal distribution of jet power, $Q_{\rm j}$,  as our input PDF, given by
\begin{equation}
    p(Q_{\rm j}) = \frac{1}{Q_{\rm j}\sigvar \sqrt{2\pi}} \exp \left[ -\frac{(\ln Q_{\rm j}/Q_0)^2}{2\sigvar^2} \right],
\end{equation}
where $\lnQ$ is the natural logarithm of the median jet power and $\sigvar$ is the standard deviation of $\lnQ$. The mean and mode of the distribution are given by $\exp(\lnQ+\sigvar^2/2)$ and $\exp(\lnQ-\sigvar^2)$ respectively. These quantities are shown for comparison in Fig.~\ref{fig:sigma}. 

As mentioned above, log-normal flux distributions are common in both disc- and jet-dominated accreting systems. In using a log-normal jet power distributions we make an implicit assumption that some form of multiplicative process imprints variability on the jet on long timescales. The mean of a log-normal distributions is larger than both the median and the mode. This asymmetry is important for UHECR production, since AGN jets must reach high powers ($\gtrsim10^{44}$\ergs) in order to accelerate CRs to energies beyond $10$EeV. Furthermore, the jets that have higher values of $\sigvar$ (more variable) also have a greater difference between the mean and median values of $Q$. 

We model the propagation of a jet with a constant mass density, $\rho_{\rm j}$. The mass density is parameterised via a density contrast with the surrounding medium, $\eta$, which is defined as the density relative to density of the ambient medium at a radius of $r=0.1~{\rm kpc}$. The jet nozzle has a constant radius, $r_{\rm j}$. We assume that the power of the jet is variable and driven by accretion rate fluctuations, and that the jet is kinetically dominated (negligible jet pressure). The kinetic power of a one-sided relativistic jet can be derived from the equations of relativistic hydrodynamics \citep[e.g.][]{taub_relativistic_1948,landau_classical_1975,wykes_1d_2019} and is given by 
\begin{equation}
    Q_{{\rm j},1} = A_{\rm j} v_{\rm j} \Gamma_{\rm j} (\Gamma_{\rm j}-1) \rho_{\rm j} c^2,
\end{equation}
where $\rho_{\rm j}$ is the jet mass density, $\Gamma_{\rm j}$ is the bulk Lorentz factor, $A_{\rm j}=\pi r_{\rm j}^2$ is the cross-sectional area of the jet nozzle and $v_{\rm j}$ is the jet velocity. We assume that the jet variability is entirely accounted for by variation of the jet velocity/Lorentz factor; thus for a given value of the two-sided jet power $Q_{\rm j}=2Q_{{\rm j},1}$ we invert the above equation to solve for $v_{\rm j}$ (or equivalently, $\Gamma_{\rm j}$). The jet is transrelativistic, so the velocity is allowed to transition between non-relativistic and relativistic regimes. The jet power ultimately determines the jet advance speed, the energy input into the lobes, the maximum CR energy and the normalisation of the particle source term (see subsequent sections). The mass input rate from the jet is $\dot{M}_{\rm j} = A_{\rm j} v_{\rm j} \Gamma_{\rm j} \rho_{\rm j}$, which, together with the rate of change of volume, determines the density in the lobes.

\begin{figure}
    \centering
	\includegraphics[width=0.9\linewidth, clip=True, trim=0.3cm 0 0 0]{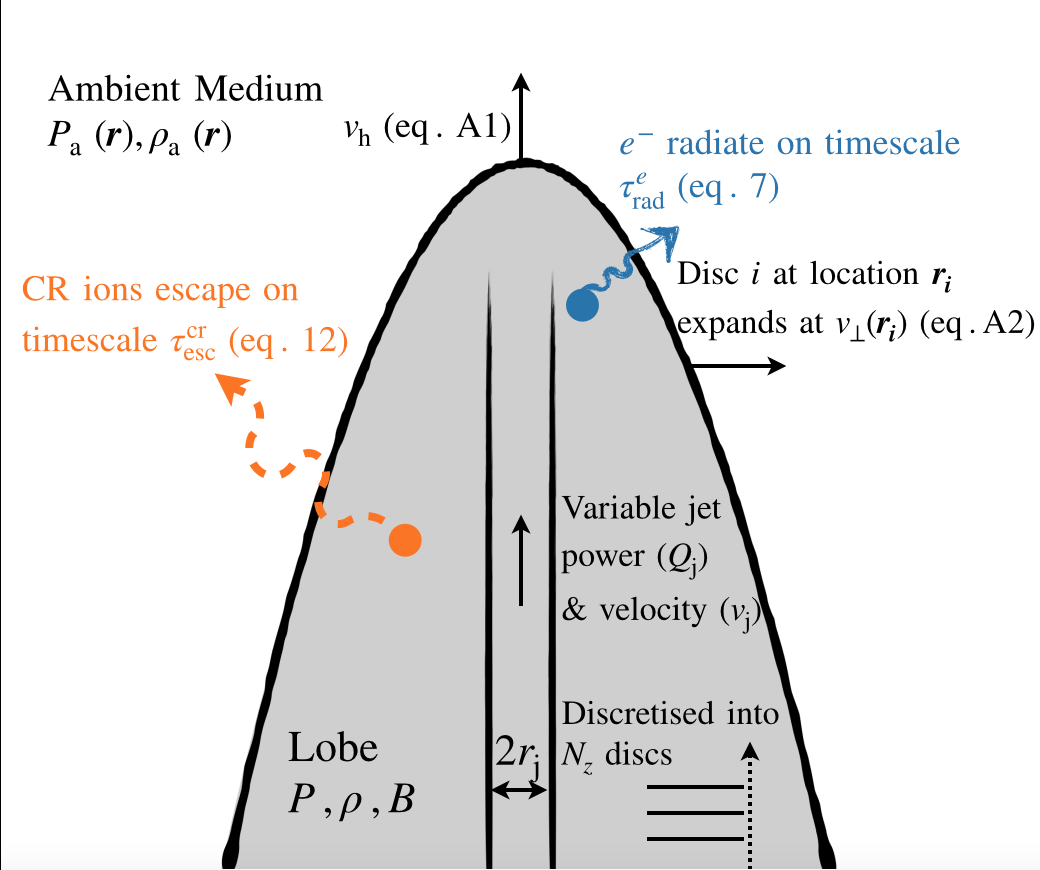}
    \caption{
    Schematic illustrating the main aspects of our method. A light, collimated jet propagates into an isothermal ambient medium with decreasing density and pressure, and inflates a cocoon or lobe. In the process it accelerates radiating electrons and CR ions, which gradually cool, and, in the case of CR ions, escape from the lobe. The dynamic model is described in Section~\ref{sec:dynamics} and Appendix A. The evolution of nonthermal particles is described in Section~\ref{sec:particles}, and the escape of CR ions is described in Section~\ref{sec:cr_escape}.
    }
    \label{fig:schematic}.
\end{figure}

\subsection{Jet and lobe dynamics}
\label{sec:dynamics}
A number of analytic and semi-analytic models for jet propagation have been developed \citep[e.g.][]{begelman_overpressured_1989,falle_self-similar_1991,kaiser_self-similar_1997,turner_raise_2018,hardcastle_simulation-based_2018}. Generally, these models concern themselves with relatively powerful jets, that are well confined and FRII-like in their morphology; our approach is similar, with the main differences being that we allow for a flickering jet power and model the expansion of the lobe by discretising into a series of cylindrical cells along the direction of propagation (the $z$-axis). 
We consider the propagation of a light jet, with typical density contrasts $\sim10^{-4}$. The speed of the jet varies depending on the power, with typical Lorentz factors ranging from non-relativistic to $\Gamma_{\rm j}\approx 10$. We assume that the advance of the jet is determined by equating, in the rest frame of the working surface, the momentum flux of the jet with that of the ambient medium \citep{marti_morphology_1997}, with an additional geometric factor (see Appendix A). We refer to the bubble inflated by the jet as a `lobe' (rather than, say, a cocoon) throughout this paper for simplicity. The sideways expansion of the lobe is set by the bow-shock jump conditions, based on the difference in pressure between the lobe and surrounding environment. The lobe pressure is calculated in a self-consistent manner from the energy injection and volume of the lobe.  

The jet propagates into an ambient medium with a density and pressure calculated from the `universal pressure profile' described by \cite{arnaud_universal_2010}. This pressure profile is characterised by a single variable, $M_{500}$, which is the mass contained within the radius where the density is $500$ times the critical density of the Universe.
The ambient medium is assumed to be isothermal, with the temperature, $T_c$, set by the $T_c-M_{500}$ relation from \cite{arnaud_structural_2005}, which gives $k_B T_c\approx2.3\,$keV for $M_{500}=10^{14}~M_\odot$. The density then follows from $\rho = \mu m_p (P/k_B T)$, where $\mu=0.62$ is the mean particle mass (we assume a fully ionized solar abundance plasma). This choice of parameters leads to a particle number density of $n \approx 2 \times 10^{-3}\,$cm$^{-3}$ at $100\,$kpc. The approximate functional dependence is a smooth broken power-law of the form $n(r) \propto (r/R_b)^{-\delta_1} \left[ 1 + (r/R_b) \right]^{(\delta_1 - \delta_2)}$, with $\delta_1=0.305$, $\delta_2=5.70$ and $R_b=715~{\rm kpc}$ for our adopted $M_{500}=10^{14}~M_\odot$.
The domain is discretised in the direction of propagation, $z$, with $N_z$ cells, and the evolution of the width of the lobe is calculated for each of the cells containing jet material. We compute a dynamic time step and a particle evolution time step, with the latter evolved during subcycles within the main dynamics loop. Our model for the jet dynamics is described further, with the relevant governing equations and more detail regarding the numerical scheme, in Appendix A. A schematic diagram describing the main features of the method is shown in Fig.~\ref{fig:schematic}.

\subsection{Energy partitioning}
\label{sec:partitioning}
To conduct our simulations, we have to decide what fraction of the jet's kinetic power is transferred to the various forms of energy. Following \cite{hardcastle_simulation-based_2018}, we assume that a fraction $\epsilon_w=0.5$ of the jet's power goes into doing $PdV$ work on the surroundings \citep[see also ][]{blundell_nature_1999,bourne_agn_2020}; the remaining half is stored as internal energy in the lobes and used to calculate the lobe pressure and sideways lobe expansion (see Appendix A). The use of this work factor is an approximation and while it is fairly accurate in estimating the overall $P dV$ work done it does not account for the fact that the work done varies over time and can in principle be higher than the energy input in periods of low jet power (see Appendix A for further details). The energetic particle populations comprise nonthermal electrons ($E_e$) and nonthermal protons and ions -- which we refer to as CRs  ($E_{c}$). We assume that a fraction $\epsilon_e$ of the jet power is transferred into nonthermal electrons, such that $dE_e/dt=\epsilon_e Q_{\rm j}$. For simplicity, we assume the same amount of energy is transferred into CRs, giving $dE_c/dt=\epsilon_c Q_{\rm j}$ where $\epsilon_c=\epsilon_e$. The equations and source terms for the particle populations are given in sections~\ref{sec:ions} and \ref{sec:electrons} and account for adiabatic losses. For the magnetic field, we set $dE_B/dt= \epsilon_w \epsilon_b Q_{\rm j}$. 

Our energy partioning factors are defined at injection, so the fact that the different components cool at different rates means that these quantities differ to the instantaneous equipartition factors often used in the literature. In addition, our quantities are defined as a fraction of the total jet power rather than relative to other components. For the equation of state of lobe plasma, we use an adiabatic index of $4/3$, although a four-fluid `effective adiabatic index' could be adopted, in a manner similar to \cite{pfrommer_simulating_2017}. The partitioning factors adopted in our simulations are given in Table~\ref{tab:fixed_params}. We experimented and chose values that gave reasonable radio luminosities and typical magnetic field strengths in the lobe of 10s of $\mu$G, in line with the results of \cite{croston_x-ray_2004}.

\subsection{Nonthermal particles}
\label{sec:particles}
AGN jets transfer energy to nonthermal particles. We model this process by evolving populations of both electrons and various CR ion species, with a source term that depends on the jet power.  We do not model the particle acceleration physics, and we remain agnostic about the details of the process. We  assume that the particles are accelerated in e.g, shocks close to the jet head, over a short acceleration time, $\tau_{\rm acc} \sim \eta_g (r_g/c)$, where $\eta_g$ is the so-called `gyrofactor \citep[e.g.][]{aharonian2000} such that $\eta_g \approx 1$ in the Bohm regime. We then allow the particles to cool in (and escape from) the lobe of the radio galaxy. Hereafter we use the subscript $e$ to denote electrons and $i$ to denote CR ion species. 

\subsubsection{Nonthermal electrons}
\label{sec:electrons}
We evolve the electrons in the lobes according to the continuity equation
\begin{equation}
\frac{dn_e(E)}{dt} = \frac{d}{dE} \left(\frac{En_e(E)}{\tau_e(E)}\right) + S_e(E, t)
\label{eq:electron_continuity}
\end{equation}
where $n_e(E)=dN_e/dE$ is the differential spectrum of electrons at energy $E$. The first term accounts for synchrotron, inverse Compton and adiabatic losses through the total cooling timescale $\tau_e(E)$, and $S_e(E, t)$ is the electron source term, given by 
\begin{equation}
    S_e(E, t) = 
    \frac{\epsilon_e Q_{\rm j}(t)}
    {\chi_e}
    E^{-p}~e^{-E/E_{\mathrm{max},e}},
\label{eq:electron_source}
\end{equation}
where $\chi_e$ is a normalisation constant and $\chi_e=\ln(E_{\mathrm{max},e}/E_{0,e})$ for $p=2$.
The injection energy is set to $E_{0,e}=10 m_e c^2$. The maximum electron energy $E_{\mathrm{max},e}$ is kept constant at 100 TeV. We have verified that this maximum energy is achievable for the range of input jet powers, under the assumption that the maximum particle energy is limited by synchrotron cooling in the jet hotspot. However, the exact value is not well constrained. We discuss the reasons for this and possible improvements further in Section~\ref{sec:discuss_params}. 
We solve equation \ref{eq:electron_continuity} following the method described by \cite{chang_practical_1970} and \cite{chiaberge_rapid_1999}, in which the equation is discretised and written in tridiagonal matrix form and then solved using a tridiagonal matrix algorithm (TDMA; also known as a Thomas algorithm).  The total radiative cooling rate due to inverse Compton and synchrotron processes for electrons with energy $E$ is
\begin{equation}
    \frac{dE}{dt}\bigg|_{\mathrm{rad}} = \frac{E}{\tau^e_{\rm rad}(E)}
    \label{eq:electron_loss}
\end{equation}
where 
\begin{equation}
\tau^e_{\rm rad}(E)=\frac{9}{4 \alpha} \frac{U_{B_{\rm crit}}}{U_{B}+U_{\rm rad}}\frac{\hbar}{E}.
\end{equation}
Here $U_B=B^2/8\pi$ is the magnetic field energy density, $U_{\rm rad}$ is the radiation field energy density and $U_{B_{\rm crit}}$ is the energy density of the Schwinger field, $B_{\rm crit}=4.41\times10^{13}$\,G. We set $U_{\mathrm{rad}}=4.17\times10^{-13}$\,erg~cm$^{-3}$, corresponding to the energy density of the cosmic microwave background (CMB) as reported by \cite{fixsen_temperature_2009}. The energy density of the CMB is equivalent to that of a $3.24\,\mu$G magnetic field. We assume that the CMB is the dominant radiation field in our simulations, which is not true close to the host galaxy, but is likely to be a good approximation once the lobe has reached $\sim10$kpc in length. Electrons and CR ions both also lose energy adiabatically as the lobe expands at the rate 
$dE/dt|_{\mathrm{ad}} = 1/3 (E/V) dV/dt$. The adiabatic loss timescale is then 
$\tau_{\mathrm{ad}} = E/(dE/dt|_{\mathrm{ad}})$, and the total cooling timescale is given by the reciprocal sum such that
\begin{equation}
\tau_e(E)= \left[ \frac{1}{\tau_{\mathrm{ad}}} + \frac{1}{\tau^e_{\mathrm{rad}}} \right]^{-1}.
\end{equation}

\subsubsection{Nonthermal ions (cosmic rays)}
\label{sec:ions}
In the same manner as the electrons, we evolve a CR distribution for each CR ionic species, $i$, with a continuity equation that differs slightly to the electrons, given by
\begin{equation}
    \frac{dn_i(E)}{dt} = S(E/Z_i, f_i, t) - \frac{n_i(E)}{\tauesc (E/Z_i)} - \frac{n_i(E)}{\tau_\mathrm{loss} (E, Z_i)} - 
\frac{n_i(E)}{\tau_{\rm ad}},
\label{eq:cr_number}
\end{equation}
where $n_i(E)=dN_i/dE$ is the differential spectrum of CR species $i$ at energy $E$, $\tauesc$ is the CR escape time (see Section~\ref{sec:cr_escape}) and $\tau_\mathrm{loss}$ is the energy loss timescale due to photopion, photodisintegration and pair production losses. The total differential CR spectrum is then $n(E,t)=\sum_i n_i(E,t)$. We do not model the coupling between the different ion species (see below). We again use a power-law with an exponential cutoff for the source term, $S(E/Z_i, f_i, t)$, given by (for $p=2$)
\begin{equation}
    S(E/Z_i, f_i, t) = 
    \frac{f_i \epsilon_c Q_{\rm j}(t)}
    {\chi_i}
    E^{-p}~e^{-E/E_{\mathrm{max,i}}}
\label{eq:cr_source}
\end{equation}
where again $\chi_i$ is a normalisation constant and $\chi_i=\ln(E_{\mathrm{max},i}/E_{0})$ for $p=2$.
The injection energy is assumed constant over time for each species and set to $E_0=10 A_i m_p c^2$ (a CR Lorentz factor of 10). The maximum energy $E_{\mathrm{max}}$ at each time step is set according to the power requirement for UHECR production based on the Hillas energy, which is discussed in the introduction, and is given by
\begin{equation}
    E_\mathrm{max,i} = 
    10~\mathrm{EeV}~Z_i \eta_{\rm H}
    \left[ 
     \frac{Q_{\rm j}}{10^{44}~\mathrm{erg~s}^{-1}} 
     \frac{\epsilon_b}{0.1}
    ~\beta_{\rm j} \right]^{1/2}.
\label{eq:max_cr_energy}
\end{equation}
where $\eta_{\rm H}=E_\mathrm{max,i}/E_H$ denotes the fraction of the Hillas energy attainable. The maximum CR energy therefore accounts for the partitioning of magnetic energy via the term $\epsilon_b$. In addition, the use of this maximum energy condition ensures that the UHECR acceleration time is always short compared to $r_{\rm j}/c$. It does also implicitly assume that particle acceleration is taking place close to the Bohm regime, but this must be the case for almost any UHECR accelerator \citep{hillas_origin_1984}. Our  method could feasibly be made more complicated by incorporating more detailed aspects of shock acceleration physics. However, our approach here captures the general behaviour of the maximum CR energy depending on time through its dependence on the square root of the jet power, which has an important impact on the results.  

The quantity $f_i$ controls the relative fraction of ion $i$ in the CR population. An enhancement of heavy ions relative to intrinsic abundance is expected on theoretical grounds  \citep[e.g.][]{ellison_monte_1981,caprioli_non-linear_2011,marcowith_microphysics_2016,matthews_particle_2020}, and there is empirical evidence for a heavy composition in both Galactic CRs \citep[e.g.][]{aglietta_cosmic_2004,blasi_diffusive_2012-1} and UHECRs beyond the ankle \citep{pierre_auger_collaboration_combined_2017,de_souza_measurements_2017}. We take a phenomonological approach for charge and mass dependent injection that approximately fits the TeV-range Galactic CR spectrum, proposed by \cite{wykes_uhecr_2017}. In this approach, the spectrum in energy per nucleon is scaled by $f_\odot Z^2/A$, where $A$ is the atomic mass number of the ion and $f_\odot$ is the solar abundance. This is equivalent to setting $f_i = f_\odot Z^2 A^{p-2}$. We note that alternative scalings for chemical enhancements have been suggested, for example by \cite{caprioli_chemical_2017}. However, the enhancement factor they propose is rather mild, scaling only as $(A/Z)^{2}$. We include all ions that are at least as abundant by number as Fe in the Sun (H, He, C, N, O, Ne, Mg, Si, Fe). 

For CR losses, we consider the normal loss mechanisms at ultrahigh energy ($\gtrsim 1$\,EeV), where protons undergo losses due to the GZK effect \citep{greisen_end_1966,zatsepin_upper_1966}, ions with $A>1$ undergo photodistintegration, and all species undergo pair production losses. For each of these processes, we tabulate rates from \textsc{crpropa} \citep{alves_batista_crpropa_2016} at $z=0$ for the CMB radiation field and the \cite{gilmore_semi-analytic_2012} model for the extragalactic background light (EBL). We only include the CMB and EBL radiation fields and we do not allow for the cascade in decreasing A and Z for heavy nuclei as they photodistintegrate and lose nucleons; instead, we treat the process as an energy loss in equation~\ref{eq:cr_number} via the $n_i/\tau_\mathrm{loss}$ term. We do not include proton synchrotron losses in our model, because, although they can be important for $10^{19}$eV protons in $\gtrsim 1$mG fields \citep[e.g.][]{aharonian2002}, the field strengths we consider instead range from $10-250~\mu$ G, such that the proton synchrotron timescale is longer than the simulation time even for the highest energy protons.

\subsection{CR Escape Time}
\label{sec:cr_escape}
The escape time is difficult to estimate. Ultra-high energy protons and ions will gradually escape out of the lobe via a combination of drifts, streaming, advection and diffusion. We expect the particles to be accelerated in shocks relatively close to the jet head, although we remain agnostic about the details of the acceleration mechanism. Particles might be advected fairly quickly into the jet lobe by hydrodynamic backflow, but this backflow slows and dissipates its energy through shocks or is broken up by vorticity and shear; as such, the flow gradually becomes slower and subsonic or transonic \citep[e.g.][]{falle_self-similar_1991,reynolds_hydrodynamics_2002,matthews_ultrahigh_2019}. The lobe is therefore broadly characterised by transonic turbulence with bulk flows of $u\sim c_s$ where $c_s$ is the sound speed. Flows can reach a significant fraction of $c$ \citep{reynolds_hydrodynamics_2002,matthews_ultrahigh_2019}, implying a minimum advective timescale  $\tau_{\rm adv} \sim 100\,{\rm kpc} / c_s \sim 1$\,Myr. However, in practice, bulk flows are slower, and less uniform, at larger distances from the jet head, and even if the CRs are advected deep into the lobe it is not clear that this facilitates quicker CR escape. This argument depends on the details of the CR escape and the exact magnetic field structure in the lobe, particularly the degree of connectivity between the lobe field and surroundings, motivating further study. Generally, we expect the UHECRs to diffuse out of the lobes more quickly than they are advected away, and even in the case of being advected long distances they must still escape the turbulent cluster magnetic field. For our purposes, we assume that the CRs escape diffusively from the lobe.

We make a rough estimate of the escape time by assuming Bohm diffusion and using the magnetic field estimate described above. If Bohm diffusion applies, the escape time for ultra-relativistic particles of energy $E$ and charge $Ze$ from a sphere of radius $L_S$ is given by 
\begin{equation}
\tauesc\ = \frac{L_S^2}{2D_B} = \frac{3 L_S^2}{2 R_g c} = \frac{3 L_S^2 Z e B}{2 E c},
\label{eq:tau_escape1}
\end{equation}
where $D_B=R_g c/3$ is the Bohm diffusion coefficient and $R_g$ is the Larmor radius.
Our lobes are not spherical, so to quickly estimate the characteristic escape distance we approximate them as ellipsoids. The centre of mass of the lobes is at the origin, since we are assuming symmetrical jet propagation. The average distance to the lobe edge, which we take as a characteristic distance of escape, can then be calculated as $L_{\mathrm{esc}} = 2 L {\cal K}(m)/\pi$. Here $L$ is the lobe length and $\cal{K}$ is the complete elliptic integral of the second kind (which ranges from 1 to $\pi/2$ in this case). $m=1-W^2/L^2$ is the square of the eccentricity and $W$ is the lobe width, defined at the widest point, which is generally close to $z=0$. Our final UHECR escape time estimate is 
\begin{equation}
\tauesc\ = 9.05~\mathrm{Myr} 
\left( \frac{L_{\mathrm{esc}}}{100~\mathrm{kpc}} \right)^2
\left( \frac{D_B}{D} \right)
\left( \frac{E/Ze}{10~EV} \right)^{-1}
\left( \frac{B}{10\mu\mathrm{G}} \right).
\label{eq:tau_escape2}
\end{equation}
We do not account for the magnetic field of the ambient medium or time delays introduced by diffusive propagation to Earth, but we discuss this and other limitations of our approach in Section~\ref{sec:discuss_escape}.

\subsection{UHECR and Radiated Luminosities}
\label{sec:method_lum}

One of our aims is to examine the effect of jet variability on the particle populations in the lobe, as well any observational consequencies for radio galaxies and UHECR origins. It is therefore useful to consider a few different (multimessenger) luminosites that are relevant for observers. We record the escaping UHECR luminosity and the radio luminosity due to synchrotron-emitting electrons, but also calculate the full broadband SED at some time steps 
The UHECR luminosity at a given time can be calculated by integrating over the particle distribution for each species, 
\begin{equation}
    L_u (>E_u) = \sum_i \int_{E_u}^\infty \frac{E n_i(E)}{\tauesc(E/Z_i)} dE
    ,
\end{equation}
or equivalently, by counting the number of CRs with $E>E_u$ that leave the system in the solution of equation~\ref{eq:cr_number}. We record the CR luminosity above  8 EeV, in order to compare to various studies of CR anisotropies, spectrum and composition by the Pierre Auger Collaboration. We also record the full CR spectrum over time.

We calculate the synchrotron power per unit solid angle, $P_\nu(\gamma_e)$, where $\gamma_e$ is the electron Lorentz factor, by assuming an isotropic pitch-angle distribution \citep[e.g][]{crusius_synchrotron_1986,ghisellini_synchrotron_1988} and a single magnetic field strength throughout the lobe volume at a given time. The total synchrotron luminosity is then calculated by integrating over the electron population to find the synchrotron emissivity per unit solid angle, $j_\nu$, and multiplying by the volume of the lobe times $4\pi$, such that the specific luminosity is given by
\begin{equation}
    L_\nu = 4 \pi j_\nu V(t) = V(t)~\int_{\gamma_0}^{\gamma_{\rm max}} n_e(\gamma) P_\nu(\gamma_e) d\gamma_e,
\end{equation}
where $\gamma_0$ and $\gamma_{\rm max}$ are the minimum and maximum electron Lorentz factors. We record $L_\nu$ at 144 MHz and 1.4 GHz for comparison with typical observation frequencies. We also present full broadband SEDs in Section~\ref{sec:sed}, calculated using \textsc{gamera} \citep{hahn_gamera_2015}. We tested our synchrotron calculation against \textsc{gamera} and the \textsc{synch} code from \cite{hardcastle_frii_1998}, finding excellent agreement.

Protons can produce gamma-rays via $pp$ and $p\gamma$ collisions. As above, we assume that the CMB and EBL radiation fields are dominant in our simulations. This assumption means that gamma-ray radiation from $p\gamma$ interactions is negligible. The energy loss timescale to $pp$ collisions is $\tau_{\mathrm{pp}} \approx 600~(10^{-4} \mathrm{cm}^{-3}/n_p)~\mathrm{Gyr}$ \citep{sikora_electron_1987}, where $n_p$ is the target proton density. The typical densities is our modelling are $n_p\lesssim 10^{-5}$cm$^{-3}$, so we expect hadronic gamma-rays to be negligible and do not consider their contribution. We also checked, using \textsc{gamera}, that the luminosity from $pp$ collisions was much lower than the IC contribution.  However, we note that hadronic gamma-rays may be important either near the base of the jet or when the lobe densities and/or lobe energy contents are higher. In addition, it may be that densities are higher outside the lobe, or inside the lobe where there is significant mixing with the external medium due to, e.g., instabilities at the contact discontinuity. Hadronic gamma-rays could therefore feasibly be detectable from CRs in these regions. 

\begin{table*}
\centering
    \begin{tabular}{c c c c}
    \hline 
     Parameter  & Value & Description & Section/Reference \\
    \hline 
    $r_{\rm j}$~(kpc)  & 0.5 & Jet radius & -- \\
    $\epsilon_e$  & 0.15 & Fraction of jet power that goes into nonthermal electrons & Section~\ref{sec:partitioning} \\
    $\epsilon_c$  & 0.15 & Fraction of jet power that goes into CR ions & Section~\ref{sec:partitioning} \\
    $\epsilon_b$  & 0.1 & Fraction of jet power that goes into magnetic field energy & Section~\ref{sec:partitioning} \\
    $\epsilon_w$ & 0.5 & Fraction of jet energy that goes into doing $pdV$ work on the surroundings & Section~\ref{sec:partitioning} \\
    $f_i(Z_i,A_i)$ & $f_\odot Z^2 A^{p-2}$ & Injection fraction for ion $i$ & \cite{wykes_uhecr_2017} \\
    $D_{\mathrm{esc}}/D_B$ & 1 & Escaping diffusion coefficient relative to Bohm & Section~\ref{sec:cr_escape} \\
    $\eta_{\rm H}$ & 0.3 & Fraction of \cite{hillas_origin_1984} energy attainable & Section~\ref{sec:ions} \\
    $Q_0$~(erg~s)$^{-1}$  & $10^{45}$ & Median jet power & Section~\ref{sec:synthetic_jet} \\
    $\sigvar$  & 1.5 & Jet variability parameter & Section~\ref{sec:synthetic_jet} \\
    $p$  & 2 & Injected particle spectral index & --\\
    $\eta$  & 10$^{-4}$ & Jet to ambient medium density contrast & -- \\
    $\rho_{\rm j}$~(g~cm$^{-3}$) & $3.5\times10^{-30}$ & Jet mass density & -- \\
    $M_{500}~(M_{\odot})$  & $10^{14}$ & Ambient medium enclosed mass (sets pressure and density profile) & \cite{arnaud_universal_2010} \\
    $k_B T_c$ (keV)  & $2.3$ & Ambient medium temperature & \cite{arnaud_structural_2005} \\
    $\alpha_{p}$ & 1 & Slope of temporal power spectrum & Section~\ref{sec:synthetic_jet} \\
    \hline
    \end{tabular}
  \caption{Main parameters used in the jet modelling, with definitions, value for the reference model, and references. 
}
  \label{tab:fixed_params}
\end{table*}


\section{Results from a single, illustrative jet model}
\label{sec:single_jet}
\begin{figure}
    \centering
	\includegraphics[width=\linewidth]{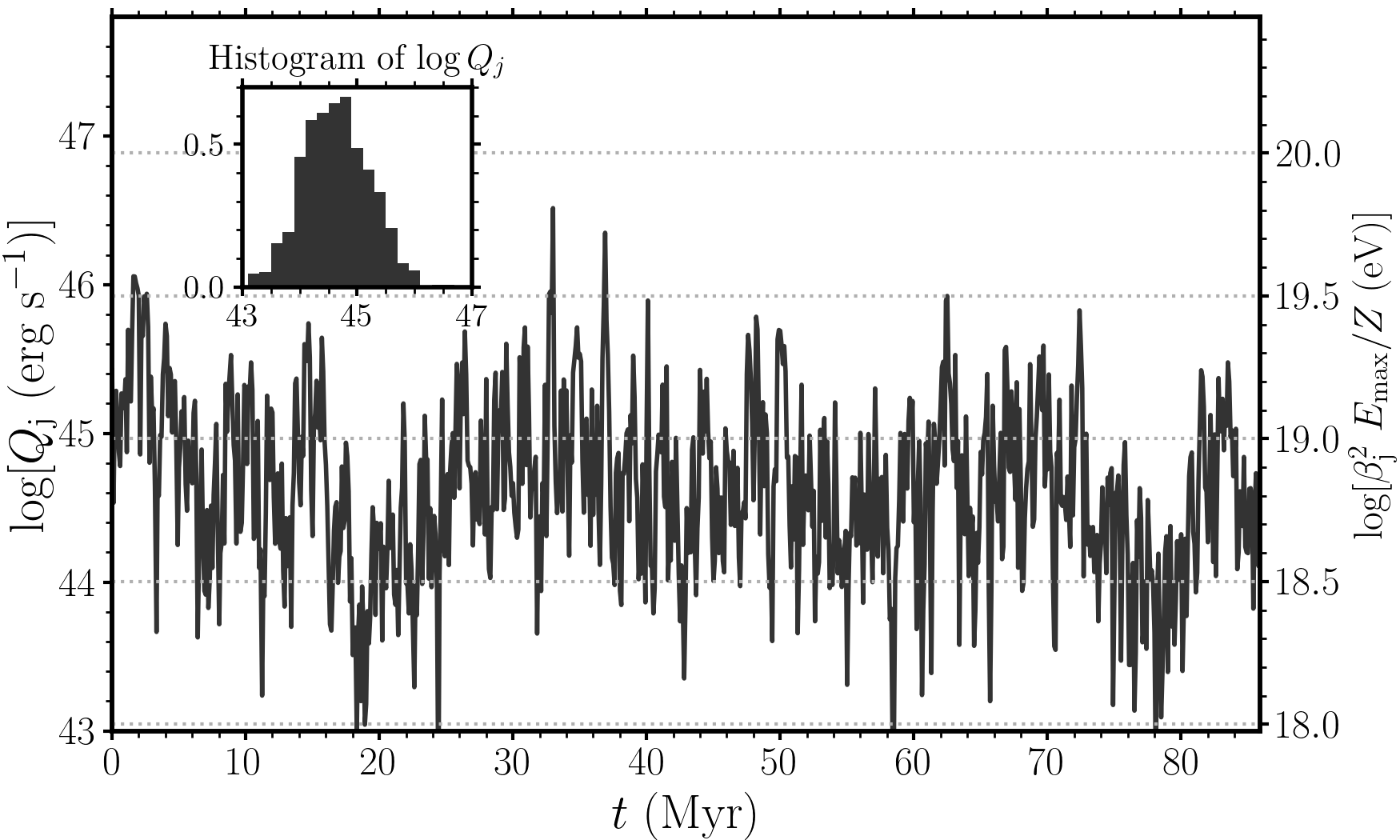}
    \caption{
    Jet power time series used for the reference model, with $Q_0=10^{45}$\ergs\ and $\sigvar=1.5$. The jet power, $Q_{\rm j}$, is shown on a log scale. For comparison,  $Q_{\rm j}$ is also shown on a linear scale in the bottom panel of Fig.~\ref{fig:evolution}. The insets show the PSD used to generate the jet power time series and the resulting histogram of jet powers (with logarithmically spaced bins). On the right-hand y-axis, the corresponding value of $\beta_{\rm j}^2 E_{\rm max}/Z$ is shown, for $\epsilon_b=0.1$ and $\eta_{\rm H}=0.3$ (the parameters used in the simulation), with dotted horizontal lines marking 0.5 dex intervals. The right-hand axis is shown to give the reader a feel for the range of maximum energies protons can attain from a jet with these parameters, in the rough range from $10^{18}$eV to $3\times10^{19}$eV.  
    }         
    \label{fig:time-series}
\end{figure}

We produce results from a single jet history. We choose to model a light ($\eta=10^{-4}$) jet with a median (two-sided) jet power of $Q_0 = 10^{45}$\,\ergs\ and a variability parameter of $\sigvar=1.5$. This median jet power is fairly typical of estimates for FRII radio galaxies, if slightly towards the lower end  \citep{godfrey_2013,ineson_2017}. We chose these values of $Q_0$ and $\sigvar$ to give reasonable overall energetics and dynamics for our source, but we also discuss the sensitivity to these parameters (as well as the PSD slope, $\alpha_p$) in Section~\ref{sec:discuss_params}. The full set of parameters for this simulation, which we refer to as our {\em reference model}, is given in Table~\ref{tab:fixed_params}. We generate the synthetic jet power time series according to the procedure in Section~\ref{sec:synthetic_jet} and evolve the jet until it is $300\,$kpc long. As the jet and lobe evolve, we record the various luminosities, lobe dimensions, lobe physical quantities and timescales, as described in the previous section. The jet power time series is shown in Fig.~\ref{fig:time-series}, with an inset showing the resulting histogram of jet powers.

\subsection{Time evolution of the system}
\begin{figure*}
    \centering
	\includegraphics[width=\linewidth]{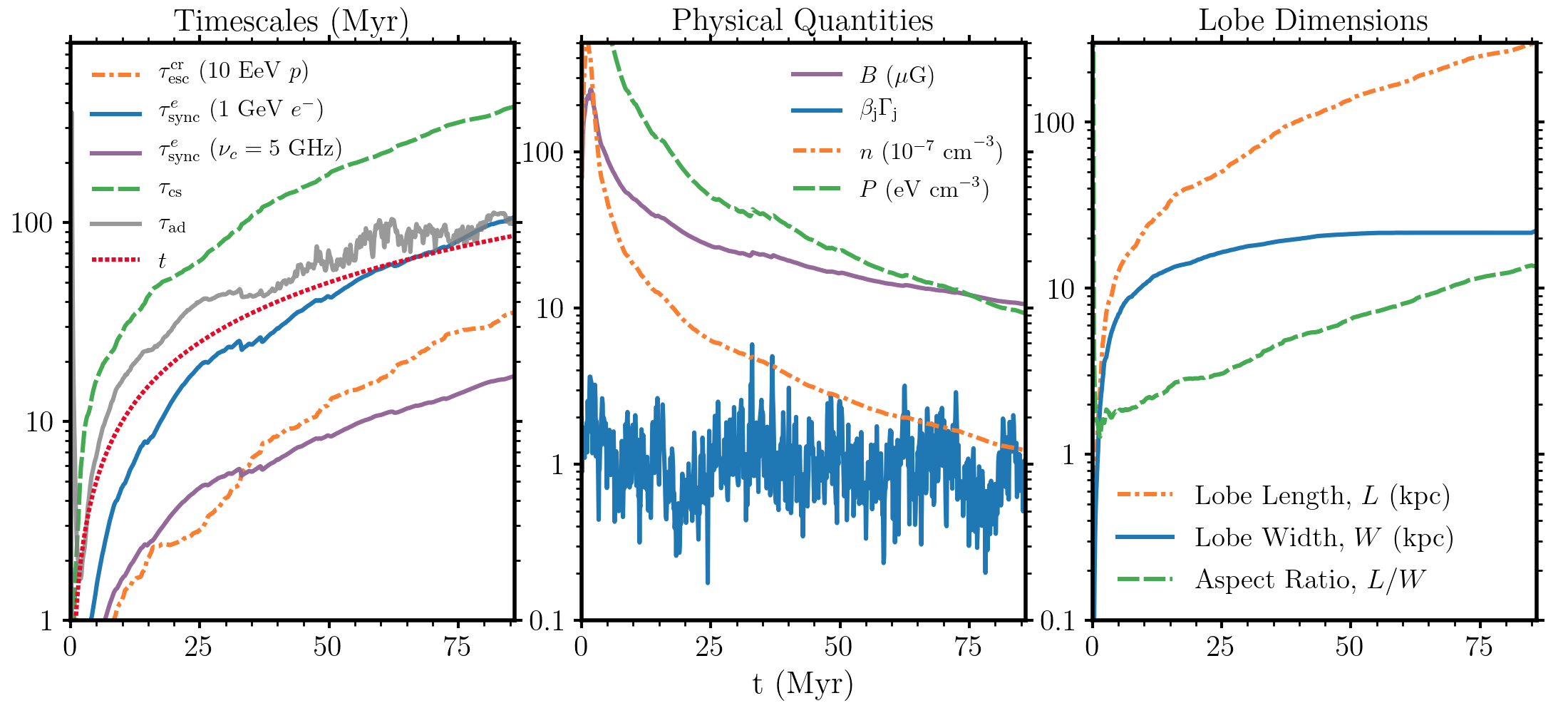}
    \caption{
    The time evolution of the timescales, physical quantities and dimensions of the lobe in the reference simulation. Each plot is shown on a logarithmic $y$ axis with the same scale on the $x$ axis. {\sl Left:} The sound-crossing time, source age, UHECR escape time, adiabatic cooling time and synchrotron cooling time as a function of time.  We give the synchrotron cooling time for $1$\,GeV electrons and for electrons emitting with a characteristic frequency of $5$\, GHz. {\sl Centre:} The lobe magnetic field strength, $B$, lobe density, $n$, lobe pressure, $P$, and jet Lorentz factor (multiplied by $\beta$) as a function of time. {\sl Right:} The length, width and aspect ratio of the radio source over time. The aspect ratio is defined as the length divided by the width.
    }         
    \label{fig:evolution}
\end{figure*}

In Fig.~\ref{fig:evolution}, we show the evolution of the important timescales, dimensions and physical quantities ($B$, $n$, $P$, $\beta_{\rm j} \Gamma_{\rm j}$) from our reference simulated radio source. The UHECR escape time and adiabatic cooling time are defined in the previous section, the synchrotron cooling time is denoted $\tau^e_{\rm sync}$, and $t$ is the age. We calculate $\tau^e_{\rm sync}$ for $1$\,GeV electrons and for electrons emitting with a characteristic frequency of $5$\,GHz, and give the CR escape time for $10$\,EeV protons. We also define the sound crossing time,  $\tau_{\rm cs}=L/c_s$ where $L$ is the lobe length and $c_s\approx700$\,km~s$^{-1}$ is the sound speed in the (isothermal) ambient medium. Our reference model has a supersonic jet and its advance is also supersonic, so the longest timescale in the system is generally $\tau_{\rm cs}$. We do not show the IC cooling time, which, for 1 GeV electrons, is approximately $1.2$~Gyr. This timescale exceeds the longest synchrotron cooling time, as expected since $U_B>U_{\rm CMB}$ is always satisfied in this case. The particle acceleration time $\tau_{\rm acc}$ is not shown since this is always less than a few hundred years.

At early times ($t\lesssim 5$\,Myr), the magnetic field is high ($\sim100\,\mu$G) before gradually dropping to around $10\,\mu$G at later times ($t\gtrsim 60$\,Myr). The strength of the magnetic field is important because it sets the hierarchy of the UHECR escape time, $\tauesc$, and electron synchrotron cooling time, $\tau^e_{\rm sync}$. Stronger (weaker) magnetic fields are better (worse) at confining UHECRs and cause quicker (slower) synchrotron cooling (see the next subsection). The UHECR escape time also depends on the size of the radio lobe. The strength of the magnetic field is proportional to the square root of the internal energy of the lobe; thus it is determined by a competition between the rate of change of volume and rate of change of energy. The former depends on the dynamics of the lobe, in particular the momentum flux, and the latter is equal to the jet power. Early on in the lobe's evolution, the environment is denser, so the relative change of volume for a given jet power is smaller. As a result, the magnetic field tends to decrease over time, because the lobe inflated by the jet finds it easier to expand. This behaviour is broadly consistent with the results of \cite{croston_particle_2018}, who find higher mid-lobe pressures (for which we expect correspondingly higher B fields) in shorter lobes. We find that the length of the lobe increases faster than the width, meaning that the lobe aspect ratio -- or axial ratio, both defined as length divided by width -- increases over time, consistent with \cite{blundell_nature_1999}.

\begin{figure}
    \centering
	\includegraphics[width=0.95\linewidth]{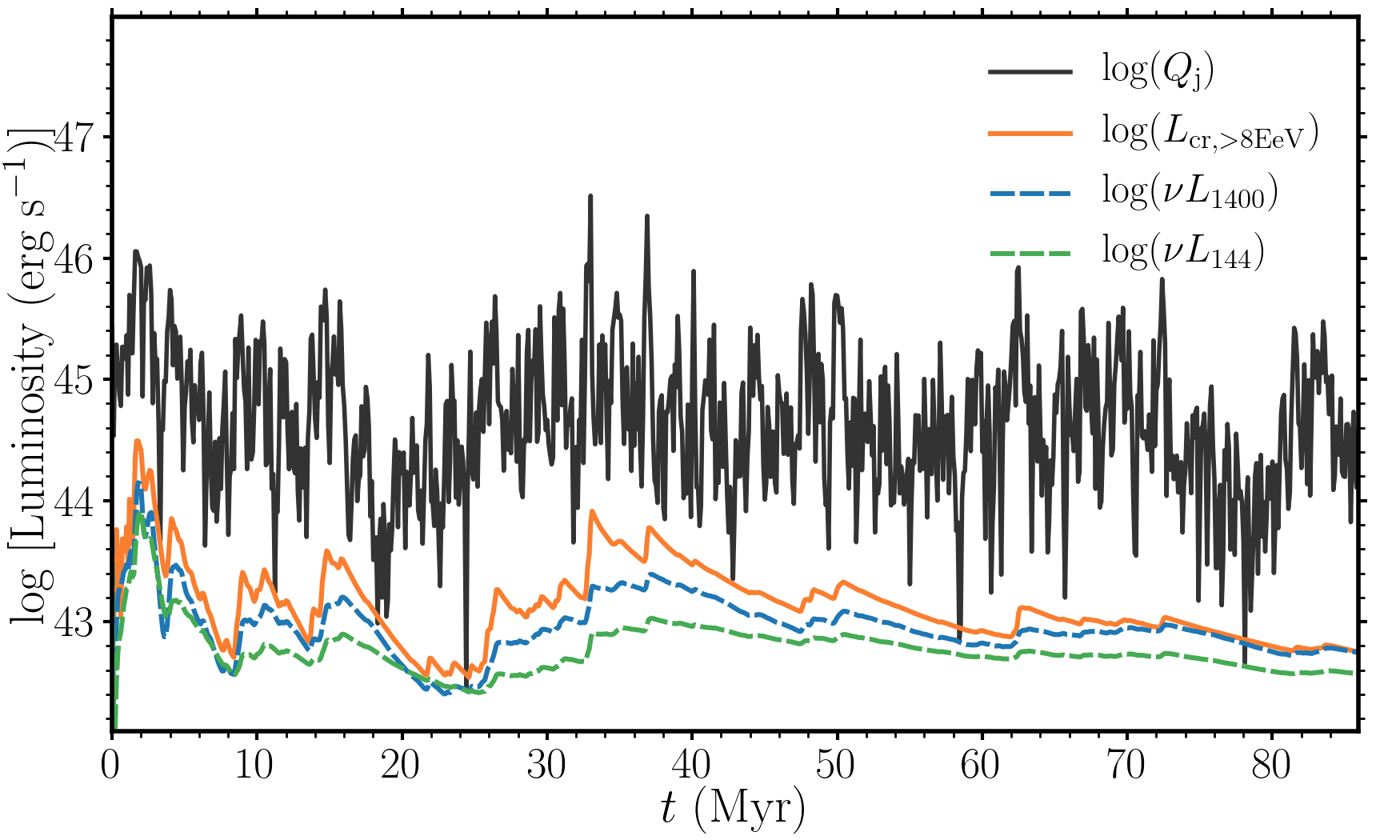}
    \caption{
    The evolution of the various luminosities over time in the reference simulation. The jet power is shown in black, matching Fig.~\ref{fig:time-series}. The UHECR luminosity is calculated above 8EeV  for comparison with studies from the Pierre Auger Observatory, and the radio luminosity is given in $\nu L_\nu$ units for observation frequencies of $144$MHz and $1400$MHz.
    }         
    \label{fig:luminosity}
\end{figure}

The jet lobe responds in a number of ways to the variability of the jet. Each time the jet power increases, the advance speed also increases, energy is injected more quickly into the lobe and the particle populations are re-energised. The exact impact on a given quantity is complicated, because it depends on the relative importance of a given period of high activity compared to the overall history of the jet up to that point. Generally, each spike in jet activity causes a corresponding increase in synchrotron and UHECR luminosity, which then decays away with a characteristic response time set by the synchrotron cooling and UHECR escape time, respectively. This can be seen in Fig.~\ref{fig:luminosity}. The effect of the cooling and escape times on the light curve is therefore to act as a `low-pass filter'; high-frequency variability is smoothed out and the power spectrum steepens above a temporal frequency $f = 2\pi / \tau$ where $\tau$ is the relevant timescale. This phenomenon has been discussed in the context of blazars by, e.g., \cite{finke_fourier_2014} and \cite{chen_particle_2016}.  

\begin{figure*}
    \centering
	\includegraphics[width=\linewidth]{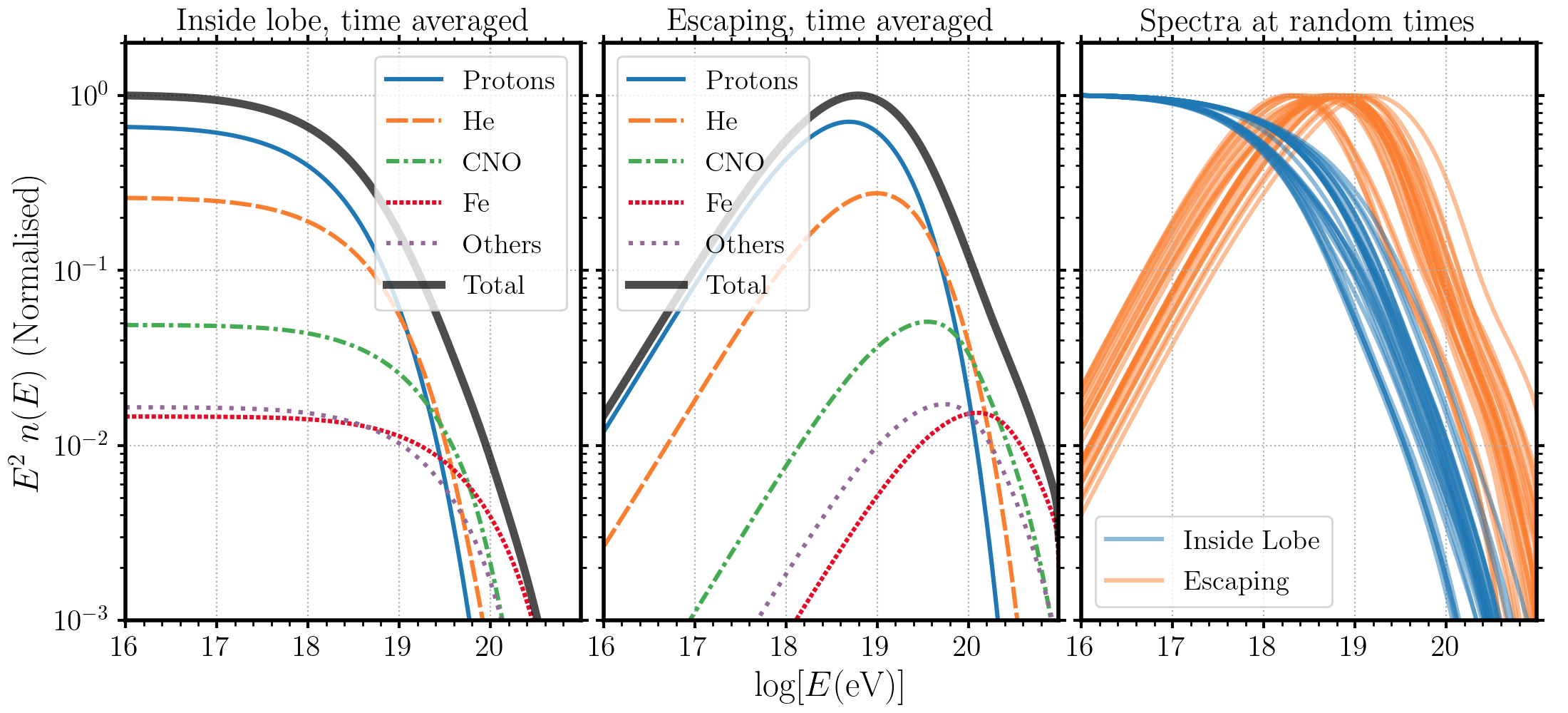}
    \caption{
    Cosmic-ray spectra from the reference simulation, in $E^2 n(E)$ units where $n(E)=dN/dE$ is the differential spectrum; an $E^{-2}$ CR spectrum appears as a horizontal line. {\sl Left:} The CR spectra inside the lobe averaged over the jet history. The thick black line shows the total CR spectrum, and the coloured lines show the contributions of individual species. The spectral shape matches the injected spectrum up to a characteristic break energy, beyond which the spectrum steepens due to the escape of high-energy CRs. {\sl Centre:} As in the left panel, but for the CRs escaping the lobe. The CR spectra behaves in an opposing manner to the left panel, with an inverted spectrum below a characteristic energy, and a flat $\sim E^{-2}$ spectrum above this energy. The spectrum steepens above a break energy determined by the range of maximum energies during the simulation due to the variable jet power; the shape of this cutoff is discussed further, for an idealised case, in Section~\ref{sec:cutoff}. {\sl Right:} CR spectra from 50 random times during the simulation. Broadly speaking, the spectra are similar to the averaged spectrum, except that jet variability causes the location of the maximum energy cutoff for the escaping CRs to jump around over-time. In addition, the value of $\tauesc$ increases over time as the lobe becomes larger, which changes the break energy
    for the CRs inside the lobe and the peak in the escaping spectrum.
    }         
    \label{fig:uhecr-spectra}
\end{figure*}

\begin{figure}
    \centering
	\includegraphics[width=\linewidth]{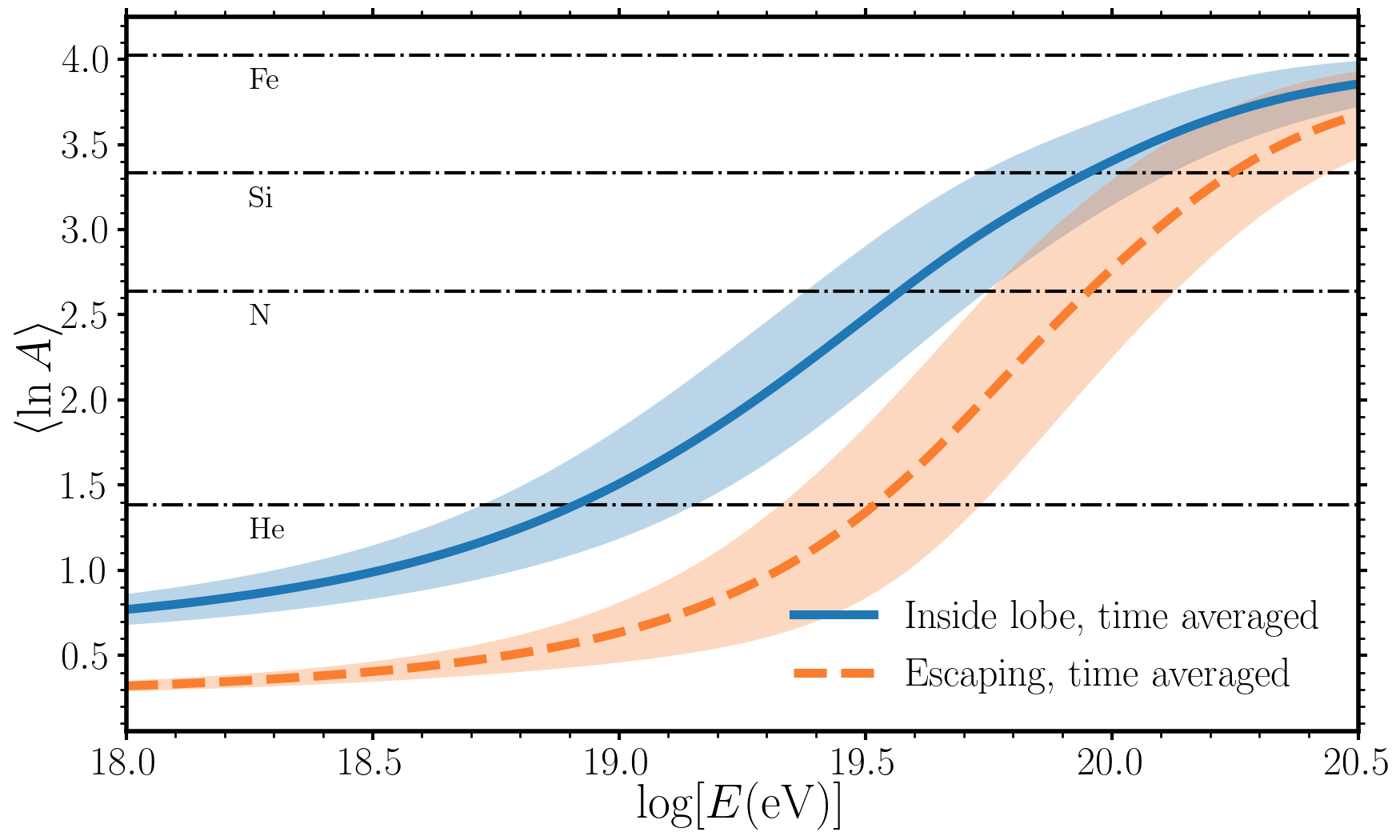}
    \caption{
    The value of $\langle\ln A\rangle$ as a function of energy in the simulation. Solid and dashed lines show the mean value across the jet history, and the shaded region shows the standard deviation at each value of $E$. The CR escape time is a rigidity ($E/Ze$) dependent quantity, so at a given energy lighter species find it easier to escape due to their lower $Z$. This means that the escaping CRs (orange) have a lighter composition than the CRs remaining inside the lobes (blue).
    }         
    \label{fig:lnA}
\end{figure}

\subsection{CR spectra and composition}
Fig.~\ref{fig:uhecr-spectra} shows a number of different representations of CR spectra, in units of $E^2 n(E)$ normalised to the maximum within the energy range $10^{16}-10^{21}$\,eV. The CRs that have not yet escaped from the lobe follow the $n(E) \propto E^{-2}$ injection spectrum up to a characteristic break energy at which the escape time equals an effective source age. In contrast, the escaping CRs follow $n(E) \propto E^{-1}$, which peaks around the same characteristic break energy. The low-energy slope for the escaping CRs, which is dominated by the protons, is a result of the assumed rigidity-dependence of the diffusion coefficient, which we have set to the Bohm diffusion coefficient such that $D_{\rm esc} \propto E/Z$. Both the internal and escaping spectra have a cutoff at higher energies. The rightmost panel of the Fig.~\ref{fig:uhecr-spectra} shows CR spectra from 30 random times throughout the simulation. There is significant diversity in the CR shape, particularly at high energies where the CRs can escape easily and are particularly sensitive to the recent activity of the jet. Below the spectral break, both the escaping and internal spectra behave in the same way as the time-averaged spectrum. 

The CR are split into different species, and a clear trend of heavier composition with increasing energy can be seen in both the internal and escaping CR spectrum. To illustrate this further, we show the value of $\langle\ln A\rangle$, the mean of the natural logarithm of the atomic mass number contributing to a given energy bin, in Fig.~\ref{fig:lnA}. Both CR acceleration and CR escape are rigidity-dependent rather than energy-dependent; as a result the value of $\langle\ln A\rangle$ naturally increases with energy. In the case of the CRs internal to the lobe, higher energies will exceed the CR energy at which light species have already escaped the lobe. In the case of the escaping CR, higher energies will exceed the maximum CR energy for light species. There is therefore a subtle difference between the increasing value of $\langle\ln A\rangle$ for the escaping and internal CRs. The fact that the escaping CRs have a lighter composition can be understood in two ways. Firstly, that at a given $E$, CRs with lower $Z$ have a higher rigidity and so find it easier to escape. Alternatively, one can think of the energy difference between the curves in Fig.~\ref{fig:lnA} as encoding the difference between the characteristic maximum energy, and the characteristic energy of escaping CRs. The difference between these is maximised when $ E_{\mathrm{max}}/Z \rightarrow \infty$ and minimised when the source lifetime is comparable to the escape time for CRs of rigidity $ E_{\mathrm{max}}/Z$. 

The abundances of species outside the source environment itself is dictated by the transport properties in this external region. In the case where energy dependent transport dominates (as may be expected if particles propagate diffusively), the resulting composition will change. Specifically, there will be an ehancement of heavy species outside the source; this effect is discussed further in Section~\ref{sec:discuss_escape}.

\subsection{The CR cutoff}
\label{sec:cutoff}
In our modelling we apply an exponential cutoff with a maximum energy that is proportional to $\sqrt{Q_{\rm j}/\beta_{\rm j}}$ (equation~\ref{eq:max_cr_energy}), as derived from the Hillas condition. As the jet power varies, so does this maximum energy (see Fig.~\ref{fig:time-series}); in more powerful episodes, both the maximum energy and normalisation associated with the source term increases, as can be seen in the range of cutoff energies in the spectra at random times shown in Fig.~\ref{fig:uhecr-spectra}. To gain more insight into the problem, we consider the total integrated spectrum (both internal and escaped CRs) and neglect the $n_i/\tau_{\rm loss}$ term. In this limit, the total spectrum is the integral of the source term, i.e. $n_i(E) = \int_0^{\Delta t} S_i(E/Z_i, f_i, t)~dt$, where $\Delta t$ is the outburst time. In the limit of large $\Delta t$, and setting $\beta_{\rm j}=1$ for simplicity, it can be shown that (see Appendix~\ref{sec:appendix_cutoff})    
\begin{equation}
    n_i(E) \propto \int_{0}^\infty p(Q_{\rm j})~Q_{\rm j}~\exp \left(- \frac{E}{k_E \sqrt{Q_{\rm j}}} \right)~dQ_{\rm j},
\label{eq:ni_pdf_general}
\end{equation}
where $k_E = \eta_{\rm H} Z_i \sqrt{\epsilon_b}$ is the constant of proportionality in the maximum energy equation. We have written this using a general PDF $p(Q_{\rm j})$ so that it can be applied to other situations where the PDF of input powers takes a different form. The specific form of equation~\ref{eq:ni_pdf_general} for our adopted log-normal $p(Q_{\rm j})$ is
\begin{equation}
    n_i(E) \propto \int_{0}^\infty 
    \frac{1}{\sigvar \sqrt{2\pi}} \exp \left[ -\frac{E}{k_E\sqrt{Q_{\rm j}}}-\frac{(\ln (Q_{\rm j}/Q_0))^2}{2 \sigvar^2} \right]~dQ_{\rm j}
\label{eq:ni_pdf_specific}
\end{equation}
This integral is not analytically straightforward, but a very good approximation can be obtained by considering only the peak of the integrand (see Appendix~\ref{sec:appendix_cutoff}). The general behaviour can already be appreciated from equation~\ref{eq:ni_pdf_general}: a spread in jet powers acts to stretch out the exponential cutoff in the spectrum, such that the cutoff function is the convolution of the jet power PDF and the source term cutoff. The effect is similar in nature to the convolution of a CR source term with a luminosity function, and a comparable discussion in the context of supernova remnants is given by \cite{shibata_chemical_2010}. The fact that the source term normalisation is proportional to $Q_{\rm j}$ biases the spectrum towards higher energies, even for a symmetric PDF. This effect is even more pronounced if $p(Q_{\rm j})$ is positively skewed as is the case for a log-normal distribution. The result is that the most important episode of activity for determining the CR maximum energy cutoff is likely to be the most energetic episode that has occured within a UHECR loss-time. 

\begin{figure}
    \centering
	\includegraphics[width=\linewidth]{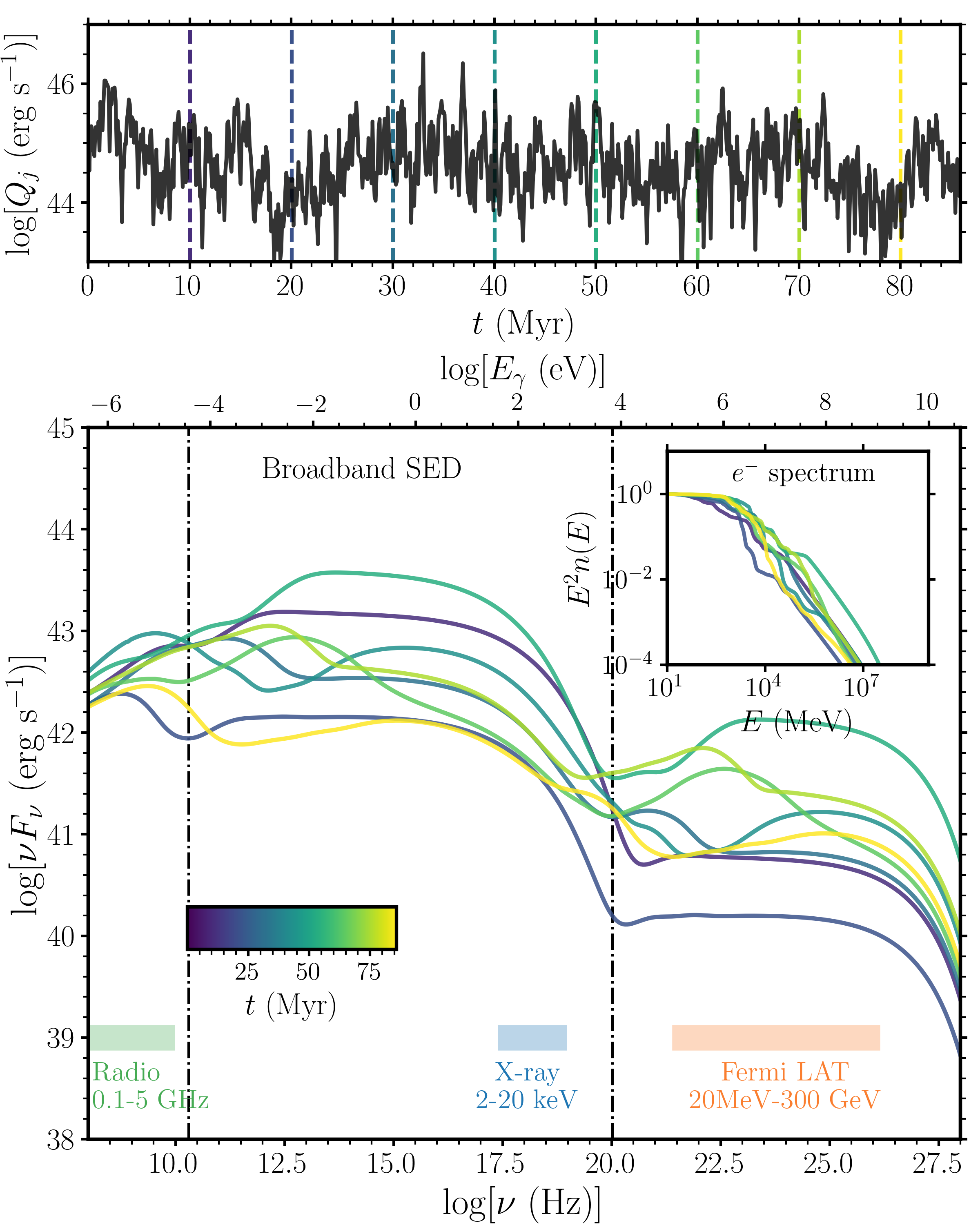}
    \caption{
   The effect of jet variability on the broadband SED. {\sl Top:} Jet power time series, as in Fig.~\ref{fig:time-series}, with 10 Myr intervals marked by dotted lines with colours corresponding to the colours used in the bottom panel. {\sl Bottom:} Broadband SED at the same 10 Myr intervals throughout the jet history, shown in $\nu F \nu$ units and with photon frequency and energy marked on separate axes. The colour-coding denotes the time in Myr at which the spectrum is calculated. Typical observational frequency ranges for radio ($0.1-5$ GHz), X-ray ($2-20$ keV) and {\sl Fermi LAT} (20 MeV-300 GeV) are marked with coloured bands. The dot-dashed vertical lines mark the frequencies at which the `proxy electrons' (electrons cooling at the same rate UHECRs are escaping) emit their synchrotron and IC radiation (see section ~\ref{sec:proxy}).
    {\sl Inset:} The corresponding electron spectrum, in the same colours and in units of $E^2 n(E)$. The dot-dashed vertical lines mark the energy of the proxy electrons discussed in Section~\ref{sec:proxy}. 
    }         
    \label{fig:sed}
\end{figure}

\subsection{Spectral Energy Distribution}
\label{sec:sed}

In Fig.~\ref{fig:sed}, we show broadband spectral energy distributions (SEDs) at 10 Myr intervals throughout the jet history. The SEDs are calculated using \textsc{gamera} as described in section \ref{sec:method_lum}. Each curve is colour-coded according to the elapsed time, and the inset axis shows the corresponding electron spectra. The top panel shows the jet power over time, as in Fig.~\ref{fig:time-series}, but with the time intervals marked with vertical dashed lines to aid interpretation of the plot. We also show characteristic observing frequencies for radio and X-ray, as well as for the {\sl Fermi Large Area Telescope} ({\sl Fermi LAT}) with coloured horizontal bands. 

The SED is characterised by a classic double-humped shape, with the low energy bump caused by synchrotron emission and the high-energy bump caused by inverse Compton scattering off the CMB. We also computed the spectrum from $pp$ collisions but found it was insignificant due to the low density of target protons in the lobes. Although a double-humped SED shape is often associated with blazars \citep[e.g][]{fossati_unifying_1998,ghisellini_theoretical_1998}, it is a generic feature of a population of electrons interacting with magnetic fields and radiation fields when the energy densities of the two fields are comparable. For IC scattering in the Thomson regime, the relative contribution of the synchrotron and IC processes is given by the ratio of the relevant energy densities ($U_B$ and $U_{\rm CMB}$ in this case). In the reference model, $U_B>U_{\rm CMB}$ at all times, and the decrease of $B$ over time causes the synchrotron bump to be more dominant at early times compared to late times.

There are clear kinks and inflection points in both the SED and electron spectrum (inset). A correspondence can be seen between features in the synchrotron hump and both the IC hump and the electron spectrum, as expected, although some of the smaller features in the electron spectrum are smoothed out in the SED since each electron energy range produces radiation over a broader range of frequencies. The kink and inflection features are caused by the variability of the jet; effectively, the result is a `many-populations' model where the electron spectrum is a series of injected populations that have each been subjected to a different cooling history. The behaviour of the electron spectrum can be understood further as follows. If a given short period of high activity dominates over the integrated historical energy input of the jet, the system behaves like a system in ``outburst'', and the spectrum resembles that of a single power-law spectrum with a cooling-break. However, in some circumstances, the period of high activity can only really be seen at high frequencies or particle energies; this is because the low energy particles haven't had time to cool and so still reflect the integrated activity, whereas the high energy particles (above the cooling break) from previous activity have all cooled, and the new powerful episode injects fresh particles in this regime that can radiate. As a result, a characteristic spectral shape is seen, with inflection points and a secondary peak and break associated with the recent ``outburst'', leading to a spectral hardening in the spectrum. A similar result is found by \citet{turner_duty-cycle_2018}, who refers to the spectral hardening as a `steep-shallow' spectrum produced by a source with short distinct outbursts. This spectral hardening effect could be an important signature of variability, so we discuss the observational perspective in Section~\ref{sec:discuss_obs}. 

\section{Discussion}
\label{sec:discussion}
\subsection{Which photon frequencies track the UHECR escape time?}
\label{sec:proxy}
From Fig.~\ref{fig:luminosity}, we can see that the radio and UHECR luminosities respond to impulses from the jet with a decay time set by the synchrotron cooling and CR escape times, respectively. We can therefore calculate at which electron energy the electrons are cooling at the same rate that the UHECRs are escaping;  we refer to the electrons at this energy as `proxy electrons' for the UHECRs. These proxy electrons emit their synchrotron and inverse Compton radiation at characeristic frequencies in radio and higher-energy bands. These radiation frequencies can be determined by comparing the UHECR escape time to the relevant cooling times. The characteristic frequency of synchrotron emission from an electron with Lorentz factor $\gamma_e$ can be conveniently written in terms of the Schwinger field such that
\begin{equation}
\nu_{\rm c}=\gamma_{e}^{2} \left(\frac{B}{B_{\rm crit}} \right) \frac{m_{e}c^{2}}{h}.
\end{equation}
Similarly, we can write the synchrotron cooling time as 
\begin{equation}
\tau_{\rm sync}^{e}= \frac{9}{4 \alpha}\left(\frac{B_{\rm crit}}{B}\right)^{2}\frac{\hbar}{\gamma_e  m_e c^2}.
\end{equation}
We can combine these two expressions and eliminate the $\gamma_e$ dependence so that
\begin{equation}
\tau_{\rm sync}^{e} \approx 41 \mathrm{Myr}~\left(\frac{B}{10\mu\mathrm{G}}\right)^{-3/2}~\left(\frac{\nu_c}{\mathrm{GHz}}\right)^{-1/2}. 
\end{equation}
Finally, equating this with equation~\ref{eq:tau_escape2} and rearranging for frequency gives
\begin{equation}
\nu_{\rm proxy} \approx 20~\mathrm{GHz}  
     \left( \frac{E/Ze}{10 EV} \right)^2 
    \left( \frac{L_{\mathrm{esc}}}{100~\mathrm{kpc}} \right)^{-4}
    \left( \frac{D_{\mathrm{esc}}}{D_B} \right)^2
    \left( \frac{B}{10~\mu\mathrm{G}} \right)^{-5}.
\label{nu_proxy}
\end{equation}
This is the characteristic frequency of synchrotron radiation emitted by the proxy electrons, which cool at the same rate that UHECRs of rigidity $E/Ze$ escape from the lobe. Unfortunately, this equation is extremely sensitive to many of the parameters, particularly the magnetic field. Decreasing the magnetic field by a factor of 10 changes $\nu_{\rm proxy}$ by a factor of $10^5$, which shifts the observing window from the radio to the far infrared ($\sim15\mu$m). The corresponding IC emission from such "proxy electrons" is,
\begin{eqnarray}
E_{\gamma}^{\rm IC} &=&\left[\frac{E_{\gamma}^{\rm CMB}}{(B/B_{\rm crit})m_{e}c^{2}}\right] h \nu_{\rm proxy}\\
&\approx& 0.44~\left(\frac{E_{\gamma}^{\rm CMB}}{6\times 10^{-4}~{\rm eV}}\right)\left(\frac{10~\mu {\rm G}}{B}\right)\left(\frac{\nu_{\rm proxy}}{20~{\rm GHz}}\right)~{\rm MeV}.
\end{eqnarray}
We have marked both $\nu_{\rm proxy}$ and $E_{\gamma}^{\rm IC}/h$ with vertical dot-dashed lines on Fig.~\ref{fig:sed}. For these specific parameters, the IC emission from the proxy electrons falls in the $\sim$MeV range, at lower energies than probed by current $\gamma$-ray telescopes such as {\sl Fermi LAT}. The similarity of the features in the SED at $\nu_{\rm proxy}$ and $E_{\gamma}^{\rm IC}/h$ is apparent -- whenever a kink or feature occurs in the synchrotron curve at $\nu_{\rm proxy}$ a matching one is seen in the IC bump, confirming that the two frequencies correspond to electrons of the same energy. 

\begin{figure}
    \centering
	\includegraphics[width=\linewidth]{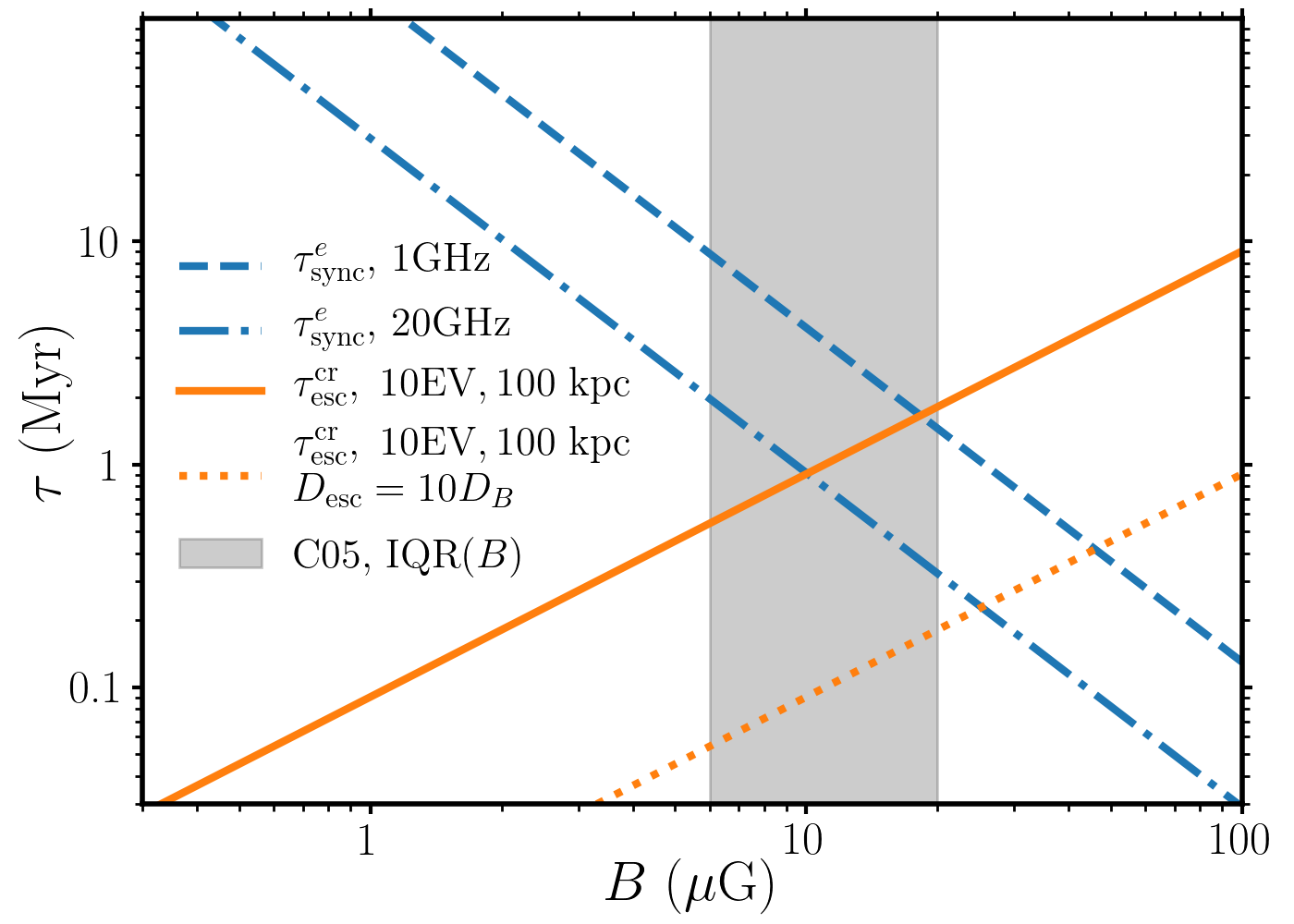}
    \caption{
    A comparison of the synchrotron cooling time, $\tau_{\rm sync}^{e}$, and CR escape time, $\tauesc$, as a function of $B$. We show $\tau_{\rm sync}^{e}$ at two different frequencies, compared to estimates of $\tauesc$ for two different escaping diffusion coefficients. $\tauesc$ is calculated for  $L_{\mathrm{esc}}=100\,$kpc and $E/Ze=10$EV. The Bohm diffusion estimate intersects with the 20GHz synchrotron curve at $B\approx10\mu G$. We also show, in a grey band, the interquartile range (IQR) of observational estimates of $B$ from \citet[][C05, their table 9]{croston_x-ray_2005}.
    }
    \label{fig:timescales}
\end{figure}

The results of this section are summarised in Fig.~\ref{fig:timescales}, where we show a comparison of synchrotron cooling times and UHECR escape times as a function of $B$ for some appropriate parameter values. To give a feel for typical magnetic fields in radio galaxy lobes, we also show the interquartile range (IQR) of observational estimates of $B$ from table 9 of \cite{croston_x-ray_2005}. Our reference model has higher magnetic fields than these observational estimates at early times but at late times the $B\sim10\,\mu$G fields fall within this range. Fig.~\ref{fig:timescales} can be thought of as a graphical representation of the simple derivation above; longer CR escape times (from larger lobes, higher $B$ or slower transport) are tracked by proxy electrons with longer synchrotron cooling times, so the value of $\nu_{\rm proxy}$ decreases as $\tau^{\rm cr}_{\rm esc}$ increases.

The above derivation is carried out for a given CR energy, but an alternative approach is to assume that the UHECRs have reached the Hillas energy, $E_H=\beta BLZe$. However, it is important to draw a distinction here between the conditions in the accelerator and the conditions when the CRs are escaping. We might expect the acceleration site to be close to the hotspot, for example in secondary shocks near the jet head \citep[e.g.][]{matthews_ultrahigh_2019}, where the characteristic velocities and magnetic field strengths are higher than the average values in the lobe. \cite{bell_cosmic_2019} showed that acceleration by multiple shocks in radio galaxy backflows provides a mechanism to reach close to the Hillas energy ($E_{\mathrm{max}} \approx 0.6 E_H)$. One advantage of these backflows is that they are not highly relativistic, thus the difficulties associated with UHECR acceleration at relativistic shocks \citep{lemoine_electromagnetic_2010,reville_maximum_2014,bell_cosmic-ray_2018} can be avoided. Let us denote the magnetic field strength, velocity and size scale of the acceleration site as $B_{\mathrm{acc}}, \beta_{\mathrm{acc}}$ and $L_{\mathrm{acc}}$, respectively. We can then use the Hillas energy to write the escape time in terms of the ratios of these quantities to the lobe values, that is
\begin{equation}
\tau_{\rm esc,H}^{\rm cr} \approx 24.5~\mathrm{Myr}~ 
\left( \frac{L_{\mathrm{esc}}/L_{\mathrm{acc}}}{50}\right)
\left( \frac{B/B_{\mathrm{acc}}}{0.1} \right)
\left( \frac{L_{\mathrm{esc}}}{100~\mathrm{kpc}} \right)
\left( \frac{\beta_{\mathrm{acc}}}{0.1} \right)^{-1}~\eta_{\rm H}^{-1},
\end{equation}
where we have used some characteristic values assuming the UHECRs are accelerated in non-relativistic ($\beta_{\rm acc} \sim0.1$) backflow shocks that are smaller than $L_{\rm esc}$ by a factor of 50 but also have stronger magnetic fields than in the lobes by a factor of 10, based on results from \cite{matthews_ultrahigh_2019}. 

\subsection{Extragalactic CR transport}
\label{sec:discuss_escape}
We have studied how CR diffusion out of the lobes affects the escaping composition, spectrum and luminosity over time, but we have not accounted for the propagation time of the CRs through extragalactic and Galactic magnetic fields, or the resulting effect on CR composition. Deflection in the extragalactic magnetic field (EGMF) introduces a time delay relative to rectilinear propagation or light travel. This delay is in addition to the escape time from the source, such that the total delay relative to the radiative signatures is $\tau_{\rm tot} = \tauesc\ + \tau_{{\rm prop}} - (R/c)$ for a source at distance $R$, where $\tau_{{\rm prop}}$ is the propagation time (outside the lobe) and we assume $R \gg L_{\rm esc}$. The propagation time depends on the diffusion coefficient as $\tau_{\rm prop} \sim R^2/D$. For the rest of this discussion, we assume that the scattering length is small compared to the source distance, $R>D/c$.  The rigidity dependence of the diffusion coefficient depends on the ratio of the Larmor radius to the coherence length of the magnetic field, $\lambda_c$ \citep[see][their eq. 1]{guedes_lang_revisiting_2020}. For low rigidity particles, $R_g < \lambda_c$, the propagation is diffusive with $D \propto R_g^{1/3} \lambda_c^{2/3}$, resulting in the expression
\begin{equation}
\tau_{\rm prop} \sim 0.7~{\rm Gyr}~
\left(\frac{E/Ze}{0.1{\rm EV}}\right)^{-\frac{1}{3}}
\left(\frac{B}{1~{\rm  nG}} \right)^{\frac{1}{3}}
\left(\frac{\lambda_c}{1~{\rm  Mpc}} \right)^{-\frac{2}{3}}
\left(\frac{R}{10{\rm  Mpc}} \right)^{2},
\label{eq:tau_prop_diffusive}
\end{equation}
where we have chosen values such that the $R_g < \lambda_C$ regime applies. This propagation time here can be extremely long, but since in a nG-strength EGMF the Larmor radius of a $1$ EV rigidity CR is $1.08$Mpc, the above regime is only realised for very large coherence lengths or particle rigidities below $1$~EV. For rigidities above $1$~EV we generally expect $R_g > \lambda_C$ in the EGMF, and we instead obtain
\begin{equation}
    \tau_{\rm prop} \sim 0.28~{\rm Gyr}
\left(\frac{\lambda_c}{1~{\rm Mpc}}\right)
\left( \frac{E/Ze}{10~{\rm EV}} \right)^{-2}
\left( \frac{R}{100~{\rm Mpc}} \right)^{2}
\left( \frac{B}{1~{\rm nG}} \right)^{2},
\label{eq:tau_prop2}
\end{equation}
where the field is assumed to be random and turbulent. For smaller $R$, such that $R<D/c$, an expression of similar form to equation~\ref{eq:tau_prop2} can be obtained in the small-angle scattering limit. The above estimates vary depending on the exact numerical prefactors adopted. 

Time delay effects have been dicussed in the context of variable UHECR sources with reference to transient sources such as gamma-ray bursts; \cite{miralda-escude_signatures_1996} show that the observed UHECR spectrum of a transient or `bursting' source tends to peak at the UHECR energy where $\tau_{{\rm prop}}$ is equal to the time since photon arrival. The basic principle is similar to the idea of `proxy electrons' we introduced in Section~\ref{sec:proxy}, except that in the case of a flickering source the particles are accelerated over many episodes rather than in a transient burst or impulse. A further time delay can be introduced by the escape from the environment surrounding the source -- for example, the cosmic rays might have to escape from a cluster. Cosmic-ray escape from cool-core clusters has been studied by \cite{kotera_propagation_2009}, with typical magnetic field strengths of $10\,\mu$G in the cluster leading to escape times comparable to those considered here. The estimates here are uncertain, but they show that the propagation times are of larger or comparable magnitude to the escape time from the lobes for CRs with $>$EV rigidities, propagating in $\sim 1~{\rm nG}$ fields for sources within $\sim10$s of Mpc. The propagation time is significantly shorter if the source distance becomes comparable to the scattering length. Any additional delay modifies the equations in Section~\ref{sec:proxy} and implies that lower energy proxy electrons may be more appropriate as tracers of the propagating UHECRs. At lower CR rigidities and for larger source distances, the propagation times quickly become restrictive with respect to typical source lifetimes. 

Diffusive propagation in the extragalactic region between the source and the Milky Way will lead to the CR density to accumulate due to the increased CR residence time in this region. Assuming the propagation is purely diffusive, once the steady-state level is reached, the CR density from a single source of distance $R$ will sit at a factor of $(Rc/D)$ larger than the CR flux level from the source. Assuming that the rigidity dependence of the diffusion takes the form $D\propto (E/Z)^{\delta}$, with $\delta >0$, rigidity dependent build up will occur. Subsequently, the steady-state density for CRs of the same energy will have heavier species abundance enhanced relative to light species, when compared to the escaping species abundance ratio from the source. 

\subsection{Observational Applications}
\label{sec:discuss_obs}
Observations of radio galaxies commonly reveal complex radio morphologies and evidence for restarting or episodic activity. In addition to these morphological signatures, our work demonstrates the potential of spectral signatures as probes of variability. In particular, inflection points where the spectrum is a sum of two or more components with different cooling breaks can be a signature of variability and a recent `outburst' (section~\ref{sec:sed}). Related behaviour is seen in Fornax A; \cite{maccagni_flickering_2020} find that the central unresolved component has a spectral break at a higher frequency than for the outer lobes, suggesting a recent injection of fresh nonthermal electrons consistent with a jet power varying on Myr timescales. Spectral hardening in gamma-rays could also be a potential signature of variability, as has been observed at GeV photon energies in Centaurus A \citep{abdalla_resolving_2020}. Variability in radio galaxies could also be important when determining the ages of the sources. There is often a discrepancy between the measured spectral age of a source (as measured from the synchrotron break frequency) and the dynamical age, known as the `spectral age problem' \citep{eilek_how_1996,harwood_spectral_2013,harwood_spectral_2015}. This problem can be alleviated if the magnetic field is below equiparition \citep{mahatma_investigating_2020}. Jet power variability can also help solve the spectral age problem in a few different ways. Firstly, a recent injection of fresh nonthermal electrons could lead to the spectral age being measured as much smaller than the true age of the source. Secondly, a series of powerful episodes can cause the source to grow faster (advance more quickly) than if the source had a steady jet with a power equal to the median or mode of the jet power history, resulting in a measured dynamical age much longer than both the true age and typical synchrotron age of the nonthermal electrons. 

Based on our modelling, we can also identify a number of interesting areas of the synchrotron and inverse Compton spectra that may provide further clues about particle acceleration and jet variability. In the gamma-rays, we suggest that characterising, in detail, the steepening of the spectrum associated with the cooling of electrons could be used to search for spectral hardening indicating recent jet activity. Observations close to the frequencies associated with proxy electrons could also be useful for predicting UHECR luminosities from individual objects although, as we have shown, the appropriate frequency for the proxy electrons is strongly dependent on the CR transport. 

Additionally, we note that our work has some implications for AGN more generally, in the context of accretion modes and radio properties of quasars. Long-term accretion variability is likely to be important for explaining why quasars with nearly identical emission line properties can have very different radio properties \citep[][Rankine et al., submitted]{richards_unification_2011}. Such a characteristic can be partly explained by a disconnect between the properties of the quasar accretion disc and the nonthermal electrons, as expected if the synchrotron cooling time is longer than the timescale over which the quasar changes its ultraviolet/optical properties. A similar principle might also be applied to the high-/low-excitation radio galaxy (HERG/LERG) dichotomy, which classifies radio galaxies based on the presence or otherwise of strong optical emission lines \citep{laing_spectrophotometry_1994,tadhunter_nature_1998,best_fundamental_2012}. The HERG/LERG distinction is thought to be fundamentally driven by Eddington fraction and accretion mode \citep{best_fundamental_2012}, but the two classes are found across a wide range of luminosities. The optical emission lines would be expected to respond on shorter timescales than the radio, so it is plausible that some HERGs and LERGs are similar objects, but with the disc caught in different states; for example, HERGs could correspond to objects that have recently (within a synchrotron cooling time) transitioned into a radiatively efficient state.  A refinement to our model would be needed to investigate this in more detail; we create a synthetic jet power time series, but one could determine the jet power from an underlying accretion model with different modes of fuelling. Future work in this area might focus on incorporating particle acceleration physics into numerical simulations that simultaneously model the fuelling of the AGN as well as the resulting jet \citep[e.g.][]{yang_how_2016,beckmann_dense_2019}. 

\subsection{Parallels with Galactic accelerators}
There are important parallels to draw between radio galaxy models for UHECRs and the phenomenology of Galactic CRs. Galactic CRs up to the knee are thought to be accelerated in the shocks of supernova remnants (SNRs), although reaching to PeV energies is a challenge and requires (at least) significant CR-driven magnetic field amplification \citep{blasi_origin_2013,bell_cosmic-ray_2013,bell_particle_2014}. Jets in X-ray binaries might also contribute, particularly at the high-energy end \citep{heinz_cosmic_2002,fender_energization_2005,cooper_high-energy_2020}, and the discovery of TeV gamma-rays from SS433 indeed suggests that X-ray binary jets might be good high-energy CR accelerators \citep{abeysekara_very-high-energy_2018}. 
Within the SNR paradigm for Galactic CRs, the most energetic galactic CRs are thought to come from young SNRs with fast-moving shocks, with the bulk of the lower energy CRs originating from older, more numerous SNRs and forming a near isotropic `soup'. As a result, it is natural that stochasticity becomes important at CR energies close to the knee \citep[e.g.][]{blasi_diffusive_2012-1,blasi_diffusive_2012-2}, because the number of sources capable of reaching these energies is relatively small. Can similar behaviour be invoked at ultrahigh energy? 
Nearby radio galaxies are compelling sources for explaining the UHECR spectrum and anisotropies at extreme ($\gtrsim 10$\,EeV) energies \citep{matthews_fornax_2018,eichmann_ultra-high-energy_2018,eichmann_summing_2019,guedes_lang_revisiting_2020}. At lower energies ($0.1$ to a few EeV) the energy loss length-scales from the GZK effect and photodistintegration are larger. Thus, the CRs in this energy range could instead come from an ensemble of background radio galaxies as discussed by \citet{eichmann_ultra-high-energy_2019} and \cite{matthews_ultrahigh_2019}, forming a roughly isotropic component with a spectrum largely determined by the integrated radio galaxy luminosity function. This isotropic component would still have to be produced in relatively local sources, since the propagation times impose a magnetic horizon at low rigidities, as can be seen from inspection of equation~\ref{eq:tau_prop_diffusive}. The overall importance of stochasticity in determining the UHECR spectrum is emphasized by the fact that the very same nearby radio galaxies that are often invoked as UHECR sources also show evidence for variable jet powers, a complex merger history and unusual radio lobe morphologies.

\subsection{Model limitations, parameter sensitivity, and future work}
\label{sec:discuss_params}
Our modelling approach is relatively simple and heuristic, and there are a number of physical characteristics of lobes and jets that are not well modelled by our scheme. We have already discussed cosmic ray transport. Another limitation is the treatment of the lobe as a single, uniform bubble, when in reality there is a complex interplay between the jet head, the jet itself and the lobe \citep[e.g.][]{falle_self-similar_1991}, and the hotspot is significantly higher pressure than the rest of the lobe. Our assumption affects the observed integrated spectrum, but also the cooling history of the electrons. A future model might include two or more `zones', with particles gradually moved from the ``hotspot'' population to the lobe at an appropriate advective rate. A technique like this has been applied in the RAiSE semi-analytic models \citep{turner_raise_2018}. Alternatively, magnetohydrodynamic simulations using tracer particles to model ensembles of electron could be used to account for the mixing of populations and non-uniformity of the magnetic field; \cite{vaidya_particle_2018} describe a recent implementation in the hydrodynamics code \textsc{pluto} \citep{mignone_pluto_2007}.

We have chosen to present results from one jet model, and the specific jet power time series chosen will clearly affect the results. Even the random number seed changes the exact jet power history, but it is more relevant to focus on the fundamental parameters of the time series: the variability parameter, $\sigvar$, the PSD slope, $\alpha_{p}$, and the median jet power, $Q_0$. Increasing the median jet power increases the typical maximum CR energy attainable in the accelerator (from equation ~\ref{eq:max_cr_energy}). As a result, the spectra shift to the right in Fig.~\ref{fig:uhecr-spectra}. The effect on the magnetic field is more nuanced. The magnetic energy being injected into the lobes increases with $Q_0$, but the energy density of the lobes also depends on the rate of change of lobe volume, and is expected to increase as $Q_0$ raised to an exponent $<1$. We therefore expect the magnetic field to be stronger in the lobes of more powerful sources, leading to slower CR escape and faster synchrotron cooling. Adopting a higher $\sigvar$ makes the jet power PDF more skewed, such that the mean jet power is higher compared to $Q_0$. However, increasing $\sigvar$ also increases the amplitude of variability, which has the effect of exacerbating variability signatures in the time evolution (Figs~\ref{fig:evolution} and \ref{fig:luminosity}) and broadband SED (Fig.~\ref{fig:sed}). Changing the shape of the PDF would also have a profound effect, with more symmetric distributions (e.g. a Gaussian) qualitatively mimicking the behaviour of a low $\sigvar$, and more asymmetric or skewed distributions increasing the difference between the mean and median $Q_{\rm j}$. Finally, we consider the PSD slope, $\alpha_p$. The pink/flicker noise PSD we used ($\alpha_p=1$) puts equal power in each logarithmic frequency range. A steeper PSD slope, $\alpha_p > 1$, leads to more power in low-frequency modes. For $\alpha_p=2$, the time series behaves as red noise and follows a random walk. As a result, once the jet enters a high-state, it tends to stay there for longer. This behaviour could in principle lead to more dramatic signatures in the UHECR luminosity or calculated spectra. The spectrum would be more sensitive to the exact location in the time series, with longer-timescale trends apparent in the luminosities and composition, and the opposite true for a flatter PSD slope. 

There are numerous other parameters that are important, such as the energy partitioning factors ($\epsilon_b$, $\epsilon_e$, $\epsilon_c$) and the particle spectral index, $p$. The particle spectral index is particularly important for the UHECR luminosity. Our chosen value of $p=2$, the canonical value for shock acceleration \citep{bell_acceleration_1978}, spreads out the total energy equally in each decade of energy, but there are many reasons to expect steepening of the CR spectrum \citep[e.g.][]{bell_cosmic_2019}. Even a mild steepening to $p=2.1$ causes (for a single specie) the fraction of total CR energy contained in UHECRs above $8$EeV to drop to $\approx2.5\%$ (compared to $\approx10\%$ for $p=2$). Finally, we note that our assumptions about the maximum particle energy could be relaxed in future work. In particular, we assumed a constant maximum electron energy, $E_{{\rm max},e}$, but it would be interesting to investigate the impact of a power-dependent $E_{{\rm max},e}$, obtained by balancing the electron acceleration time with the synchrotron time in the amplified magnetic field in the jet hotspot. The value of $E_{{\rm max},e}$ primarily affects the shape of the SED in Fig.~\ref{fig:sed}. It may also be necessary to account for more detail plasma physics for both electrons and CR ions, since, particularly in relativistic shocks, the growth rate of CR-driven turbulence at the shock can be severely limiting in terms of the maximum particle energy \citep[e.g.][]{reville_maximum_2014,araudo2016,araudo2018,bell_cosmic-ray_2018}.

\section{Conclusions}
We have presented a study of particle acceleration in radio galaxies with jet powers that vary over time according to a flicker noise power spectrum, and examined the effect of this flickering on the UHECR and electron populations accelerated by the source. Our main conclusions are as follows:
\begin{itemize}
    \item A log-normal distribution of jet powers results in a mean jet power that is significantly higher than the median, and especially, the mode (Fig.~\ref{fig:sigma}). If such a distribution is realised on long timescales ($\gtrsim$Myr) in nature, it provides a situation where UHECRs can be accelerated in the most powerful episodes, even when the most commonly observed jet power is significantly lower. This is true more generally for positively skewed probability distribution functions for the jet power and could also apply to systems that undergo distinct outbursts. 
    \item We present a semi-analytic model for studying the evolution of radio jets and lobes and the particle populations they accelerate and contain (section~\ref{sec:method}). Our model is influenced by other semi-analytic approaches \citep[e.g.][]{turner_raise_2018,hardcastle_simulation-based_2018}, but explicitly accounts for a variable power, transrelativistic jet and also models the non-radiating CR ion populations and their diffusion out of the lobe. The lobe is vertically discretised into a  series of discs and the advance of the jet head, expansion of each disc and resulting lobe properties are calculated self-consistently in an iterative manner, also feeding into the particle solver. This method involves a series of assumptions, and various limitations of the method together with possible improvements are discussed in Section~\ref{sec:discussion}. We present one single jet model with a flickering jet power and use it to illustrate some general principles relating to the problem of particle acceleration in a variable AGN jet source. 
    \item We find that the synchrotron and UHECR luminosities track the jet power, but with a characteristic response that is determined by the radiative cooling time and UHECR escape time, respectively (Fig.~\ref{fig:luminosity}). The cooling and escape times act like a low-pass filter on the jet power time series, smoothing out fast variability. The response time changes throughout the jet history, as, typically, the lobe size increases and the magnetic field decreases over time. 
    \item The variable jet power creates a series of electron populations with different normalisations, which are in turn each subjected to a different cooling history. Clear kinks and inflection points are observed in the electron spectrum and resultant SED (Fig.~\ref{fig:sed}). Jet variability therefore produces complexity and curvature in the electron and radiated spectrum that clearly differs from the expected spectrum for constant jet power, or for a single electron population.
    \item The CR spectrum inside the lobe follows the spectrum at acceleration, but with a break energy determined by the CR escape time (Fig.~\ref{fig:uhecr-spectra}). This break is smoothed out due to the varying magnetic field and lobe size. For our assumed energy-dependence of the diffusion coefficient, the escaping spectrum follows an $E^{-p+1}$ spectrum up to a cutoff at a maximum CR energy dictated by the jet power.
    \item The total integrated CR spectrum (the sum of the escaping and internal spectrum) displays a cutoff that is similar to a stretched exponential. In the absence of CR losses, the shape of this exponential is equal to the expectation value of the cutoff function, such that the differential spectrum obeys (see Section~\ref{sec:cutoff} and Appendix~B)
    $$
    n(E) \propto \int_{0}^\infty 
    \exp \left[-\frac{E}{k_E}\sqrt{\frac{1}{Q_{\rm j}}} \right]
    Q_{\rm j}~p(Q_{\rm j})~dQ_{\rm j}.
    $$
    In situations where the maximum particle energy is determined by the source power, it may be possible to search for evidence of jet variability by studying the cutoff shape. In addition, the cutoff is likely in general to be smoother and more gradual than a pure exponential, and the energy of the cutoff is more reflective of the most powerful outbursts than the mean or median maximum energy. 
    \item We introduce the idea of `proxy electrons' -- nonthermal electrons that are cooling at the same rate that UHECRs are escaping from the source (section~\ref{sec:proxy}). We use this concept to derive an optimum observing frequency, the frequency of radiation that best tracks the UHECR luminosity. For our assumptions about CR escape, this frequency is typically $20$\,GHz for a 10\,$\mu$G field, with corresponding inverse Compton emission at $\sim$MeV energies. These values suggest radio luminosity may be reasonable UHECR luminosity proxy, but this is strongly dependent on physical parameters in the individual sources as well as the uncertain CR transport physics. 
    \item Some local radio galaxies (such as Cen A and Fornax A) display characteristics of variable jet powers in a morphological sense. These same radio galaxies make for compelling UHECR sources. We discussed parallels with supernova remnants and propose a scenario somewhat analogous to Galactic CR production, in which the highest energy CRs are accelerated in recent outbursts in local sources, whereas UHECRs around the ankle would come from a near-isotropic background of radio galaxies as discussed by, e.g., \cite{eichmann_ultra-high-energy_2019}.
\end{itemize}
Our work highlights the influence of variability in the jet power on particle acceleration in radio galaxies and their resulting spectra and morphologies. It also emphasizes the importance of stochasticity for UHECR source models, particularly for local sources and at the highest energies.

\section*{Acknowledgements}
We thank the anonymous referee for a thorough, helpful and constructive report. JM acknowleges a Herchel Smith research fellowship at Cambridge. We gratefully acknowledge the use of the following software packages: astropy \citep{the_astropy_collaboration_astropy_2013,the_astropy_collaboration_astropy_2018}, matplotlib 2.0.0 \citep{matplotlib}, Gamera \citep{hahn_gamera_2015}. We would like to thank Tony Bell, Katherine Blundell, Sam Connolly, Will Alston and the attendees of the UK radio-loud AGN meetings for useful discussions. 

\section*{Data Availability}
The data produced as part of this work are available from the authors on reasonable request.



\bibliographystyle{mnras}


\appendix

\section{Dynamic model for the jet and lobe}
We consider a light jet propagating in the $z$ direction into an isothermal ambient medium with a decreasing density and pressure. Our jet model involves the discretisation of the $z$-domain into a series of disc-shaped cells with height $\Delta z = 0.01\,$kpc. The length of the jet after $n$ time steps is then $L_n = n \Delta z$ where the dynamic time-step $\Delta t = \Delta z / v_h$ (note that there is a separate time step for the particle solver, see Section~\ref{sec:particle_solver} of this appendix). The advance speed $v_h$ of the jet head is chiefly governed by ram pressure balance, such that $v_h$ depends on the jet speed and density constrast with the surrounding medium and is given by \citep[e.g.][]{marti_morphology_1997}
\begin{equation}
    v_h = \frac{dz_{\rm j}}{dt} = \xi \frac{\sqrt{\eta^*_r}}{1+\sqrt{\eta^*_r}} v_{\rm j},
\end{equation}
where $\eta^*_r$ follows the notation of \cite{marti_morphology_1997} and represents the relativistic generalisation of the density contrast. For nonrelativistic internal energies in the jet beam and ambient medium, as assumed here, the specific enthalpy can be neglected and $\eta^*_r = \Gamma_{\rm j}^2 \rho_{\rm j} / \rho_a(z)$, where $\rho_a(z)$ is the ambient medium density at distance $z$. Note that $\eta^*_r$ tends to the density ratio $\rho_{\rm j}/\rho_a$ as $\Gamma_{\rm j}\rightarrow 1$. $\xi=1/4$ is a geometric factor, which we chose to give reasonable advance speeds of $\approx 0.01c$. The physical motivation for the factor $\xi$ is that the jet termination shock has a smaller area than the bow shock, so the pressure acts over a wider area and slows the advance. 

\begin{figure}
    \centering
	\includegraphics[width=\linewidth]{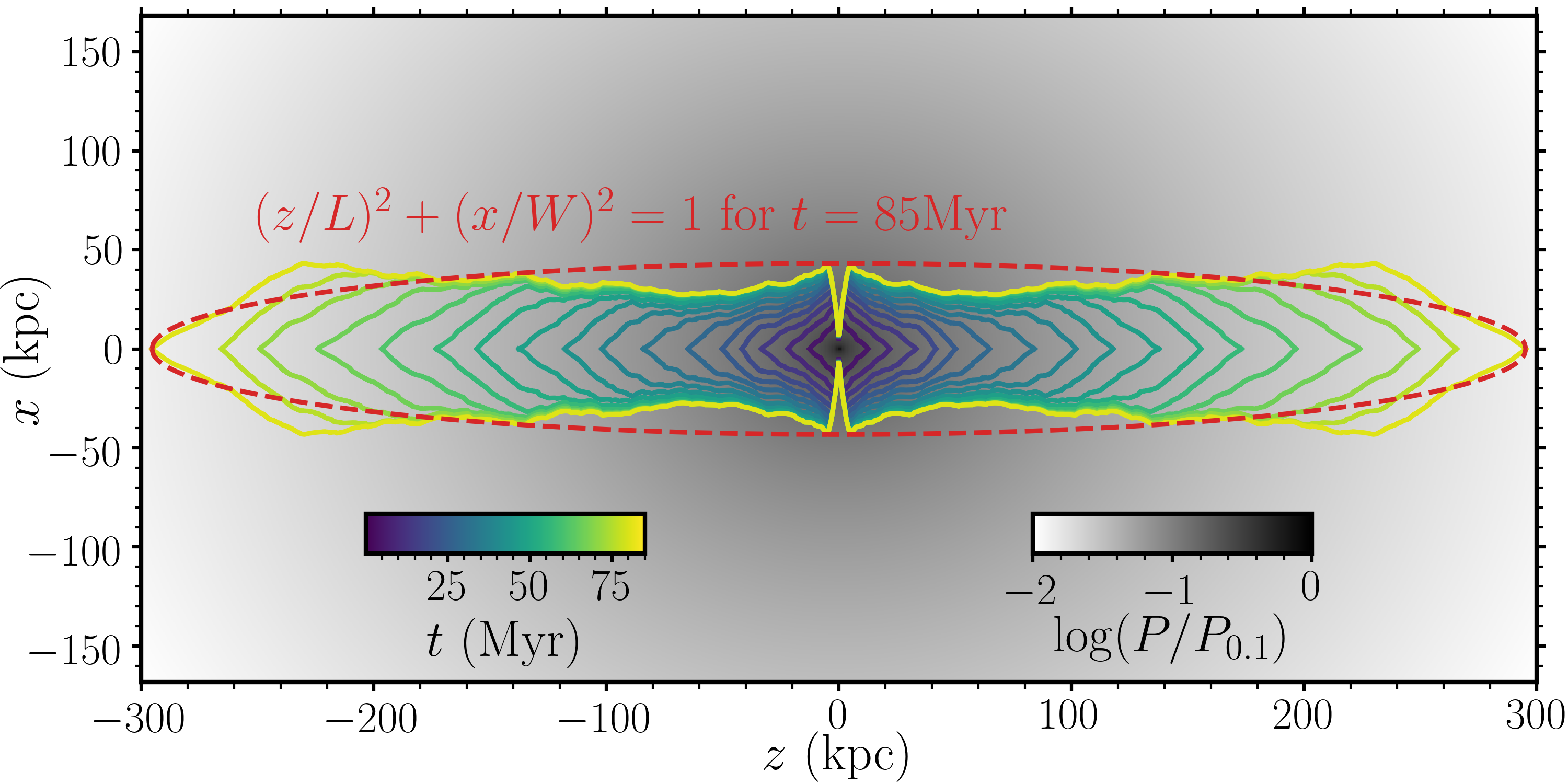}
    \caption{
    The evolution of the jet lobe in the reference model over time. The coloured lines show the outline of the lobe at $5$\,Myr intervals, with the colour scale matching that in Fig.~\ref{fig:sed}. The red line shows the outline of the ellipse used to calculated the escape time in equation~\ref{eq:tau_escape2} near the end of the source evolution at $t=85$\,Myr. The background colour shows the pressure of the surrounding environment, normalised to the pressure at $0.1$ kpc from the centre ($P_{0.1}=76.3$~eV~cm${-3}$).  
    }
    \label{fig:shell}
\end{figure}

Each disc-shaped cell has zero width until the jet material enters it, at which point the width of cell $i$ is evolved according to $w_{n+1,i} = w_{n,i} + v_\perp \Delta t$.  This sideways expansion  is governed by the expansion speed of the bow-shock. If the shocked gas between the bow-shock and the contact discontinuity is in pressure equilibrium with the lobe, then, from momentum conservation, we can write
\begin{equation}
    v_\perp(\boldsymbol{r}) = \frac{P - P_a(\boldsymbol{r})}{\rho_a(\boldsymbol{r})} \sin \phi,
\end{equation}
where $\phi$ is the angle between the local bow shock normal and the direction of jet propagation and $\boldsymbol{r}$ is the position of the edge of the cell. This equation states that the transverse expansion of the lobe is driven by the pressure difference between the lobe and the ambient medium. $v_\perp$ is a function of position since $P_a$ and $\rho$ both decrease away from the origin. The ambient pressure is modelled according to the pressure profile described in Section~\ref{sec:dynamics}. We obtain $P$ by calculating the internal energy in the lobes and converting to pressure
\begin{equation}
    P = \frac{\gamma - 1}{V(t)} \int_0^t \epsilon_w Q_{\rm j}(t) dt,
\end{equation}
where $\gamma=4/3$ is the adiabatic index and the volume is obtained by integrating over each cylindrical cell of width $w$, that is
\begin{equation}
    V(t) = \int_0^{L(t)} 2 \pi w(z,t) dz. 
    \label{eq:volume}
\end{equation}
This set of equations is evolved numerically as an initial value problem and thus requires an initial condition for the starting lobe geometry - for this purpose, we assume that the base of the lobe initially has a width $L(t_1)/2$. The outline of the lobe at $5$\,Myr intervals is shown in Fig.~\ref{fig:shell}. 

\subsection{Particle Solver and Time-stepping}
\label{sec:particle_solver}
The particle solver operates inside the main dynamic loop and undergoes a series of subcycles per dynamic time step. The time step is calculated as 
\begin{equation}
    \Delta t|_p = C_f \frac{N_m}{dN_m/dt} \bigg|_{\mathrm{min,~all~m}},
\end{equation}
where $C_f=0.4$ is a Courant-type number and $m$ is an index denoting the energy bin, and we only include bins with $N_m > 10^{-15}~N_{m,\mathrm{max}}$ so as to avoid small time-steps from fast-cooling bins with negligible numbers of particles. The TDMA solver and synchrotron emissivity calculation form part of a Python package called \textsc{msynchro}. We have tested the TDMA code using single injections of powerlaw and delta function electron distributions and comparing to analytic formulae. We have also tested the synchrotron code against the \textsc{synch} code from \cite{hardcastle_frii_1998} and the \textsc{gamera} code \citep{hahn_gamera_2015}. The tests result in extremely close agreement and are documented within the \textsc{msynchro} code repository\footnote{The \textsc{msynchro} code is publicly available at \url{https://github.com/jhmatthews/msynchro}}.

\section{The shape of the cutoff in the cosmic ray spectrum}
\label{sec:appendix_cutoff}
\begin{figure}
    \centering
	\includegraphics[width=\linewidth, clip=true, trim=0 0.2cm 0 0.2cm]{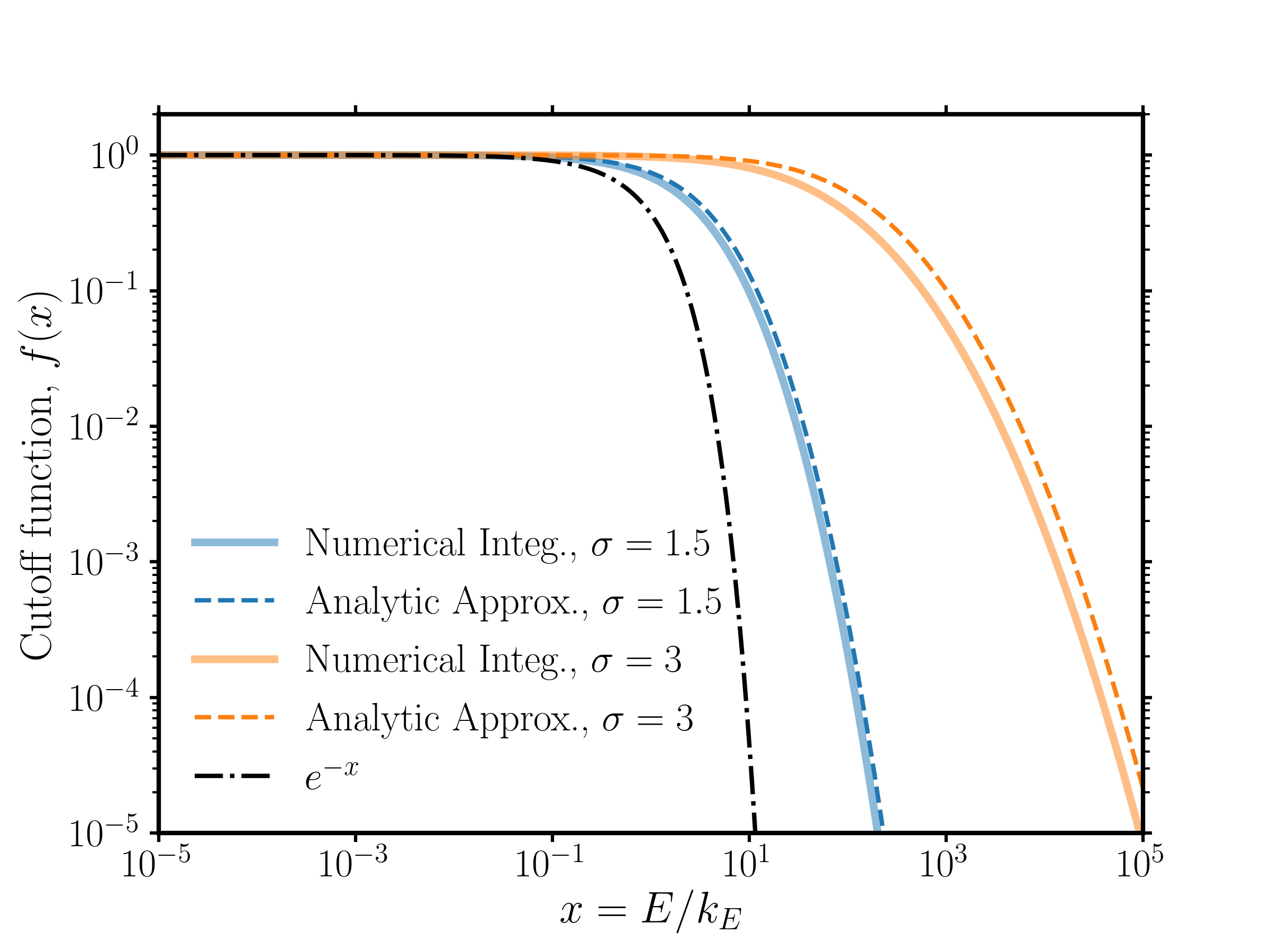}
    \caption{
    A comparison of the approximation to the cutoff function, $f(x)$ (equation~\ref{fx_approx}), to a numerical integral of equation~\ref{fx_integral}, for two values of $\sigvar$. In both cases we set $Q_0=1$. We also show a pure exponential on the same plot, showing that more variable sources create cutoffs that are more `stretched out', and with a higher characteristic maximum energy (see alsoSection~\ref{sec:cutoff}) than for a steady source with $Q_{\rm j}=Q_0$. 
    }
    \label{fig:fx_approx}
\end{figure}

The variation in jet power causes a corresponding variation in the maximum CR energy. The maximum CR energy is given by equation~\ref{eq:max_cr_energy}, and can be written more simply as $E_\mathrm{max,i} = k_E \sqrt{Q_{\rm j}/\beta_{\rm j}}$ where $k_E = \eta_{\rm H} Z_i \sqrt{\epsilon_b}$. The variation in maximum energy means that the source term has a cutoff function that varies over time, with a corresponding effect on the overall spectrum. To illustrate this, we consider the continuity equation~\ref{eq:cr_number} governing the evolution of CR ions. We focus on one CR
specie (protons) and ignore both the cooling and escape loss terms, such that $dn_i(E)/dt = S(E/Z_i, f_i, t)$. The source term $S(E/Z_i, f_i, t)$ is given by equation~\ref{eq:cr_source}, and, if we neglect the (weak) variation in the ${\ln(E_\mathrm{max}/E_0)}$ term, we can write it as 
\begin{equation}
    S(E/Z_i, f_i, t) \propto 
    Q_{\rm j}(t)~E^{-p}~e^{-E/E_{\mathrm{max,i}}}.
\end{equation}
We can integrate $dn_i(E)/dt$ so that, after a time interval $\Delta t$ the spectrum is
\begin{equation}
    n_i(E) \propto  
    E^{-p}~
    \int_0^{\Delta t} 
    Q_{\rm j}(t)~\exp \left[-\frac{E}{k_E}\sqrt{\frac{\beta_{\rm j}}{Q_{\rm j}(t)}} \right] dt.
\end{equation}
If we take the limit of large $\Delta t$, we can make use of the ``law of the unconcious statistician'' \citep[e.g.][]{degroot2012}, which states that the expected value of a function $g({\cal Z})$, where ${\cal Z}$ is a continuous random variable with PDF $p({\cal Z})$, is given by the integral of $p({\cal Z})g({\cal Z})$ over all ${\cal Z}$. Thus, if we set $\beta_{\rm j}=1$ for simplicity, the spectrum obeys the proportionality
\begin{equation}
    n_i(E) \propto \int_{0}^\infty p(Q_{\rm j})~Q_{\rm j}~\exp \left(-\frac{E}{k_E \sqrt{Q_{\rm j}}} \right)~dQ_{\rm j},
\end{equation}
as given in Section~\ref{sec:cutoff}, and where the positive integration domain reflects the fact that $Q_j$ cannot be negative. We now focus on the specific case of our adopted log-normal form for $p(Q_{\rm j})$, which gives 
\begin{equation}
    n_i(E) \propto \int_{0}^{\infty} 
    \exp \left[-\frac{E}{k_E}\sqrt{\frac{1}{Q_{\rm j}}} -\frac{(\ln (Q_{\rm j}/Q_0))^2}{2\sigvar^2} \right]~dQ_{\rm j}
\end{equation}
If we make a change of variables, defining $x=E/k_E$ and $y=Q_{\rm j}/Q_0$, we can represent the integral in the form
\begin{eqnarray}
f(x,Q_0,\sigma)=\int_{-\infty}^{\infty}  y \exp\left[-\frac{x}{\sqrt{y Q_0}}-\frac{(\ln y)^{2}}{2\sigvar^{2}} \right] d\ln y,
\label{fx_integral}
\end{eqnarray}
where $f(x,Q_0,\sigma)$ represents the integral in the previous expression, such that $n_i(E) \propto f(x,Q_0,\sigma)$.
We approximate the integral over $\ln y$ by assuming it is dominated by the peak of the integrand term. The peak in the integrand occurs at $x=2y_{p}^{1/2}(\ln y_{p}/\sigma^{2}-1)$. Inverting this relation gives 
\begin{eqnarray}
y_{p}~(x, \sigma)=\left(\frac{\sigvar^{2}x/4}{W(e^{-\sigvar^{2}/2}\sigvar^{2}x/4)}\right)^{2}
\end{eqnarray} 
where $W(z)$ is the Lambert W-function, which satisfies the relation $W(z)\exp[W(z)]=z$.
Adopting this integrand peak $y$-value, $y_{p}$, the functional form for the cutoff is found,
\begin{eqnarray}
f(x,Q_0,\sigma)\approx y_{p}\exp\left[-\frac{x}{\sqrt{Q_0 y_p}}-\frac{(\ln y_p)^{2}}{2\sigvar^{2}} \right].
\label{fx_approx}
\end{eqnarray} 
This description is found to accurately characterise the cutoff shape, over ten decades in $f$ and for a range of values of $\sigvar$. To illustrate this, we show a comparison between the analytic approximation (equation~\ref{fx_approx}) with a numerical solution to equation~\ref{fx_integral} in Fig.~\ref{fig:fx_approx}, for two values of $\sigvar$.  

\bsp	
\label{lastpage}
\end{document}